\documentclass[longauth]{aa}  
\usepackage{orcidlink}
\usepackage{graphicx}
\usepackage{txfonts}

\hypersetup{colorlinks=true,linkcolor=black,citecolor=blue,filecolor=blue,urlcolor=blue}

\begin{document}

   \title{The breaking of cold traps and onset of titanium in ultra-hot Jupiter atmospheres: WASP-189b in context}

    \author{
    S.\ Pelletier\inst{1,*}\orcidlink{0000-0002-8573-805X},
    R.\ Allart\inst{2}\orcidlink{0000-0002-1199-9759},
    V.\ Vaulato\inst{1}\orcidlink{0000-0001-7329-3471},
    D.\ Ehrenreich\inst{1,3},
    E.\ Cristo\inst{4},
    L.\ Dauplaise\inst{2},
    C.\ Lovis\inst{1}\orcidlink{0000-0001-7120-5837},
    L.\ Moranta\inst{2,5}\orcidlink{0000-0001-7171-5538},
    N.\ B.\ Cowan\inst{6,7}\orcidlink{0000-0001-6129-5699},
    T.\ Forveille\inst{8}\orcidlink{0000-0003-0536-4607},
    \'E.\ Artigau\inst{2,9}\orcidlink{0000-0003-3506-5667},
    F.\ Baron\inst{2,9}\orcidlink{0000-0002-5074-1128},
    S.\ C.\ C.\ Barros\inst{4,10}\orcidlink{0000-0003-2434-3625},
    B.\ Benneke\inst{11,2},
    X.\ Bonfils\inst{8},
    F.\ Bouchy\inst{1}\orcidlink{0000-0002-7613-393X},
    M.\ Bryan\inst{12},
    B.\ L.\ Canto Martins\inst{13}\orcidlink{0000-0001-5578-7400},
    R.\ Cloutier\inst{14}\orcidlink{0000-0001-5383-9393},
    N.\ J.\ Cook\inst{2}\orcidlink{0000-0003-4166-4121},
    J.\ R.\ De Medeiros\inst{13}\orcidlink{0000-0001-8218-1586},
    X.\ Delfosse\inst{8}\orcidlink{0000-0001-5099-7978},
    R.\ Doyon\inst{2,9},
    J.\ I.\ Gonz\'alez Hern\'andez\inst{15,16}\orcidlink{0000-0002-0264-7356},
    D.\ Lafreni\`ere\inst{2}\orcidlink{0000-0002-6780-4252},
    I.\ de Castro Le\~ao\inst{13}\orcidlink{0000-0001-5845-947X},
    L.\ Malo\inst{2,9},
    C.\ Melo\inst{17},
    L.\ Mignon\inst{1,8},
    C.\ Mordasini\inst{18}\orcidlink{0000-0002-1013-2811},
    F.\ Pepe\inst{1}\orcidlink{0000-0002-9815-773X},
    R.\ Rebolo\inst{15,16,19}\orcidlink{0000-0003-3767-7085},
    J.\ Rowe\inst{20},
    N.\ C.\ Santos\inst{4,10}\orcidlink{0000-0003-4422-2919},
    D.\ S\'egransan\inst{1},
    A.\ Su\'arez Mascare\~no\inst{15,16},
    S.\ Udry\inst{1}\orcidlink{0000-0001-7576-6236},
    D.\ Valencia\inst{12},
    G.\ Wade\inst{21,22},
    C.\ Cadieux\inst{1,2,},
    L.\ Dang\inst{23}\orcidlink{0000-0003-4987-6591},
    E.\ Deibert\inst{23},
    X.\ Dumusque\inst{1}\orcidlink{0000-0002-9332-2011},
    D.\ O.\ Fontinele\inst{13},
    Y.\ G.\ C.\ Frensch\inst{4}\orcidlink{0000-0003-4009-0330},
    F.\ Genest\inst{2}\orcidlink{0000-0003-0602-9106},
    R.\ de Lima Gomes\inst{2,13}\orcidlink{0000-0002-2023-7641},
    N.\ Gromek\inst{14}\orcidlink{0009-0000-1424-7694},
    M.\ J.\ Hobson\inst{1}\orcidlink{0000-0002-5945-7975},
    V.\ Krishnamurthy\inst{6},
    K.\ Al Moulla\inst{4,1}\orcidlink{0000-0002-3212-5778},
    A.\ Osborn\inst{8,14,24}\orcidlink{0000-0002-5899-7750},
    C.\ Piaulet-Ghorayeb\inst{2,25},
    B.\ N.\ Skinner\inst{14,26,}\orcidlink{0009-0000-9731-2462},
    A.\ Srivastava\inst{2}\orcidlink{0009-0009-7136-1528},
    A.\ K.\ Stefanov\inst{15,16}\orcidlink{0000-0002-6059-1178},
    J.\ P.\ Wardenier\inst{18,2}\orcidlink{0000-0003-3191-2486},
    D.\ Weisserman\inst{14},
    V.\ Yariv\inst{8}\orcidlink{0009-0005-2775-1589}
    }

    \institute{
    \inst{1}Observatoire de Gen\`eve, D\'epartement d’Astronomie, Universit\'e de Gen\`eve, Chemin Pegasi 51, 1290 Versoix, Switzerland\\
    \inst{2}Institut Trottier de recherche sur les exoplan\`etes, D\'epartement de Physique, Universit\'e de Montr\'eal, Montr\'eal, Qu\'ebec, Canada\\
    \inst{3}Centre Vie dans l’Univers, Facult\'e des sciences de l’Universit\'e de Gen\`eve, Quai Ernest-Ansermet 30, 1205 Geneva, Switzerland\\
    \inst{4}Instituto de Astrof\'isica e Ci\^encias do Espa\c{c}o, Universidade do Porto, CAUP, Rua das Estrelas, 4150-762 Porto, Portugal\\
    \inst{5}Plan\'etarium de Montr\'eal, Espace pour la Vie, 4801 av. Pierre-de Coubertin, Montr\'eal, Qu\'ebec, Canada\\
    \inst{6}Department of Physics, McGill University, 3600 rue University, Montr\'eal, QC, H3A 2T8, Canada\\
    \inst{7}Department of Earth \& Planetary Sciences, McGill University, 3450 rue University, Montr\'eal, QC, H3A 0E8, Canada\\
    \inst{8}Univ. Grenoble Alpes, CNRS, IPAG, F-38000 Grenoble, France\\
    \inst{9}Observatoire du Mont-M\'egantic, Qu\'ebec, Canada\\
    \inst{10}Departamento de F\'isica e Astronomia, Faculdade de Ci\^encias, Universidade do Porto, Rua do Campo Alegre, 4169-007 Porto, Portugal\\
    \inst{11}Department of Earth, Planetary, and Space Sciences, University of California, Los Angeles, CA 90095, USA\\
    \inst{12}Department of Physics, University of Toronto, Toronto, ON M5S 3H4, Canada\\
    \inst{13}Departamento de F\'isica Te\'orica e Experimental, Universidade Federal do Rio Grande do Norte, Campus Universit\'ario, Natal, RN, 59072-970, Brazil\\
    \inst{14}Department of Physics \& Astronomy, McMaster University, 1280 Main St W, Hamilton, ON, L8S 4L8, Canada\\
    \inst{15}Instituto de Astrof\'isica de Canarias (IAC), Calle V\'ia L\'actea s/n, 38205 La Laguna, Tenerife, Spain\\
    \inst{16}Departamento de Astrof\'isica, Universidad de La Laguna (ULL), 38206 La Laguna, Tenerife, Spain\\
    \inst{17}European Southern Observatory (ESO), Karl-Schwarzschild-Str. 2, 85748 Garching bei M\"unchen, Germany\\
    \inst{18}Space Research and Planetary Sciences, Physics Institute, University of Bern, Gesellschaftsstrasse 6, 3012 Bern, Switzerland\\
    \inst{19}Consejo Superior de Investigaciones Cient\'ificas (CSIC), E-28006 Madrid, Spain\\
    \inst{20}Bishop's University, Dept of Physics and Astronomy, Johnson-104E, 2600 College Street, Sherbrooke, QC, Canada, J1M 1Z7, Canada\\
    \inst{21}Department of Physics, Engineering Physics, and Astronomy, Queen’s University, 99 University Avenue, Kingston, ON K7L 3N6, Canada\\
    \inst{22}Department of Physics and Space Science, Royal Military College of Canada, 13 General Crerar Cres., Kingston, ON K7P 2M3, Canada\\
    \inst{23}Department of Physics and Astronomy, University of Waterloo, 200 University W, Waterloo, ON N2L 3G1, Canada\\
    \inst{24}Department of Physics, The University of Warwick, Gibbet Hill Road, Coventry, CV4 7AL, UK\\
    \inst{25}Department of Astronomy \& Astrophysics, University of Chicago, 5640 South Ellis Avenue, Chicago, IL 60637, USA\\
    \inst{26}Origins Institute, McMaster University, 1280 Main St W, Hamilton, ON, L8S 4L8, Canada\\
    \inst{*}\email{stefan.pelletier@unige.ch}
    }
    
   \date{Received 16 March 2026; accepted 29 May 2026}

  \abstract
   {
   Condensates are ubiquitous to all Solar System planets with significant ($>$10\,mbar) atmospheres. The same is true for most exoplanets with characterised atmospheres, with even ultra-hot exoplanets being able to form clouds on their cooler nightsides. One high-interest condensate is titanium, a highly refractory element that as an oxide (TiO) is a potent optical absorber long believed to be a driver of thermal inversions in highly irradiated exoplanet atmospheres.  Observations have shown that some ultra-hot Jupiters have strong thermal inversions despite being significantly Ti-depleted, likely due to cold trapping, raising doubts about whether TiO is the main cause of their inverted temperature structures.
   }
   {
   Our aim was to retrieve the titanium-to-iron ratio of the dayside atmosphere of the ultra-hot Jupiter WASP-189b to determine whether the full titanium budget is accounted for in the gas phase.
   }
   {
   We analysed thermal emission observations of WASP-189b taken with the HARPS and NIRPS spectrographs using different atmospheric retrieval prescriptions to infer the planet's atmospheric thermal structure and composition.
   }
   {
   We observed Fe and Ti signals in cross-correlation and measured a sub-solar Ti/Fe ratio for WASP-189b using both free and chemical equilibrium retrieval approaches. We find the Ti/Fe of the planetary atmosphere to be between 28\% and 81\% (1-$\sigma$ bounds) that of the stellar value.  
   }
   {
   The slight underabundance of Ti with respect to Fe on the dayside atmosphere of WASP-189b suggests that some titanium is missing from the gas phase, potentially due to a partial nightside cold trap. In the context of the ultra-hot Jupiter population, the onset of titanium in exoplanetary atmospheres appears to occur progressively, coinciding not with when thermal inversions begin but rather when the vapourisation threshold of titanium is reached on the nightside. 
   }


    \authorrunning{Pelletier et al.}
    \titlerunning{NIRPS dayside of WASP-189b}

   \maketitle
   \nolinenumbers

\section{Introduction}

Despite being a trace species in the stratosphere, ozone (O$_3$) plays an outsized role in shaping Earth's atmospheric thermal structure. Owing to their high opacity at ultraviolet wavelengths, O$_3$ molecules can efficiently prevent energetic solar photons from reaching the surface~\citep{malicet_ozone_1995}. Heat can therefore be deposited at pressure layers in the atmosphere where O$_3$ is present, causing temperature to increase (rather than decrease as expected from radiative cooling) with altitude. This O$_3$-driven stratospheric heating has profound implications regarding Earth’s atmospheric circulation and climate~\citep[e.g.][]{kasting_long-term_1989}.

For short-period exoplanets, which experience irradiation levels that are hundreds to thousands times that of Earth, the same phenomenon of having potent short-wavelength absorbers present in the upper atmosphere can produce full blown thermal inversions. In particular, titanium is of interest in the context of exoplanetary atmospheres as its oxidised form, titanium oxide (TiO), has long been predicted to be the primary driver of thermal inversions in highly irradiated planetary atmospheres due to its high optical cross-section and relatively high natural abundance in the Universe~\citep{hubeny_possible_2003, fortney_unified_2008}. However, while inverted thermal profiles have since systematically been detected on the hottest known giant exoplanets~\citep[e.g.][]{petz_pepsi_2025}, whether TiO is the driving agent responsible for these temperature inversions is not always clear. Indeed, TiO has remained a notoriously challenging molecule to unambiguously detect in exoplanet atmospheres. Historically, the main challenges were (and to some degree remain) the limited spectral resolution and wavelength coverage of space-based facilities in the optical, allowing resolution of only a few broadband features, plus inaccuracies in line lists hampering detections from ground-based high-resolution spectrographs~\citep[e.g.][]{hoeijmakers_search_2015, evans_detection_2016}. 

While models predict that TiO should dominate the optical spectrum of hot and ultra-hot Jupiter atmospheres, as it does for late-type stars of comparable temperatures~\citep{kirkpatrick_standard_1991}, observations at low or medium spectral resolving powers ($R < 5000$) tend to show more attenuated features that generally cannot be unambiguously associated with TiO~\citep[e.g.][]{sing_hst_2013, evans_detection_2016, mikal-evans_emission_2019, edwards_ares_2020}. In cases where observations show no clear features, 
analyses made with atmospheric retrievals missing important absorbers such as H$^{-}$ (bound-free) or of data with remaining systematics can erroneously infer bounded abundance constraints for TiO and VO, as these combined form a pseudo-continuum that can replicate larger-than-expected transit depths at bluer wavelengths~\citep{arcangeli_h-_2018, edwards_ares_2020}. 
Similarly, the combination of TiO and a stronger inversion can compensate for the omission of reflected light in eclipse observations of ultra-hot Jupiters~\citep{pelletier_enriched_2026}.

Observations at high resolution ($\lambda / \Delta \lambda = R  > 25\,000$) resolve unique patterns of up to thousands of individual spectral lines rather than a few broad molecular bands that may overlap with features from other absorbers. Although inaccuracies in previously available TiO cross-sections were known to be substantial in certain wavelength regions~\citep{hoeijmakers_search_2015}, the current most widely used line list~\citep{mckemmish_exomol_2019, mckemmish_hybrid_2024} is at least good enough for repeatable detections on one exoplanet~\citep{prinoth_titanium_2022, prinoth_time-resolved_2023}. Nevertheless, searches for TiO at high resolution have often yielded non-detections~\citep{merritt_non-detection_2020, hoeijmakers_hot_2020, hoeijmakers_mantis_2024, pelletier_vanadium_2023} or shown signals with large velocity offsets~\citep{nugroho_high-resolution_2017, serindag_is_2021, cont_detection_2021}, often raising doubts regarding the line list reliability. It is possible that using an imperfect opacities has resulted in TiO signals being teased out in only the most favourable conditions and has produced ambiguous non-detections otherwise. In particular, if line positions are more accurate at some wavelengths than others, a cross-correlation function over the full bandpass of an instrument will not necessarily add constructively to the overall signal~\citep{hoeijmakers_search_2015, prinoth_titanium_2022}.

One way to circumvent difficulties relating to detecting TiO is to indirectly infer its presence by searching for atomic Ti instead~\citep{hoeijmakers_hot_2020}. Unlike molecular TiO, which has complex rotational and vibrational modes, line transitions from atomic Ti are relatively well characterised~\citep{kurucz_including_2017} and regularly used to infer stellar abundances~\citep[e.g.][]{chavez_isotopic_2009}. Furthermore, neutral atomic Ti should at least be partly present in tandem with TiO at observable pressures under most temperature conditions above the condensation threshold ($\sim$2000 -- 3500\,K), based on equilibrium chemistry predictions~\citep{kitzmann_fastchem_2024}. Exceptions include extremely high temperature conditions, in which neither TiO or Ti will be present due to near total dissociation and ionisation, such as the case of KELT-9b~\citep{hoeijmakers_atomic_2018, kasper_confirmation_2021}.  Despite this, ultra-hot Jupiters showing strong signatures of other metals have often shown absent or weaker than expected Ti signals~\citep[e.g.][]{hoeijmakers_hot_2020, hoeijmakers_mantis_2024, pelletier_vanadium_2023, prinoth_titanium_2025}.

One hypothesis that has been put forth to explain the apparent lack of strong Ti or TiO features in hot and ultra-hot Jupiter atmospheres is a cold-trap mechanism~\citep{evans_optical_2018, hoeijmakers_hot_2020}. As Ti (T$_{\mathrm{cond}}$\,$\sim$\,1600\,K) is more refractory (i.e.\ has a higher condensation temperature) than most other metals, such as Fe, Mg, or Si (T$_{\mathrm{cond}}$\,$\sim$\,1350--1400\,K), it will be removed from the gas phase of a cooling parcel of gas before other less refractory elements~\citep{lodders_solar_2003}. More specifically, Ti is expected to condense to a liquid or solid form via TiO$_2$, which can then form perovskite (CaTiO$_3$) or other calcium titanites~\citep{lodders_titanium_2002}. The extreme ($>$2000\,K) temperatures generally probed at photospheric pressures on the dayside and terminator regions of ultra-hot Jupiter atmospheres are likely too elevated for any significant cloud mass to form~\citep{parmentier_thermal_2018, helling_cloud_2021}. However, rainout could still occur deeper in the atmosphere at pressures below the dayside thermal inversion (vertical cold trapping) or on the colder nightside hemisphere (horizontal, or nightside, cold trapping), depending on the heat redistribution efficiency~\citep{spiegel_can_2009, parmentier_3d_2013}. Compared to TiO molecules, TiO$_2$ is expected to have a lower energy barrier for self-condensation~\citep{jeong_electronic_2000}, which can then form seed particles that enable more efficient heterogeneous nucleation~\citep{fegley_atmospheric_1996, gao_aerosol_2020}. In the absence of rainout, condensates formed in an air parcel on the nightside would recirculate to the hot dayside and vaporise back to the gas phase. However, larger cloud particles tend to sink to deeper atmospheric layers, which, depending on the gravitational settling and vertical mixing timescales, can drive a depletion of more refractory species such as Ti from the observable upper atmosphere~\citep{powell_formation_2018, powell_transit_2019}.  

Whether or not a cold trap depletes select elements of a given refractoriness can significantly affect the observed spectrum of a planet.  As different metals have varying levels of refractoriness, hot Jupiter atmospheres may show condensation sequences wherein elements are sequentially removed from the gas phase of the observable atmosphere via condensation and rainout with decreasing temperature~\citep{lothringer_uv_2020, lothringer_uv_2022, roman_clouds_2021, pelletier_vanadium_2023}. A notable example of this is the dominant signature of Fe at optical wavelengths routinely observed in T$_{\mathrm{eq}}$ $\gtrsim$ 2200\,K ultra-hot Jupiters~\citep[e.g.][]{pino_neutral_2020, nugroho_searching_2020, nugroho_detection_2020, ehrenreich_nightside_2020}, compared to its apparent absence in colder T$_{\mathrm{eq}}$ $\lesssim$ 2100\,K hot Jupiters~\citep[e.g.][]{sedaghati_spectral_2021, stangret_high-resolution_2022, bello-arufe_transmission_2023, sicilia_gaps_2024}. Another is the measured depletion or underabundance of Ti-species in planets on the colder end of the ultra-hot Jupiter spectrum (T$_{\mathrm{eq}}$ $\sim$ 2200 -- 2350\,K), such as WASP-76b, KELT-20b, MASCARA-4b, and WASP-121b~\citep{kasper_unifying_2022, johnson_pepsi_2023, gandhi_retrieval_2023, pelletier_vanadium_2023, pelletier_enriched_2026, prinoth_titanium_2025}, compared to the strong presence of Ti-bearing species in the hotter (T$_{\mathrm{eq}}$ $\gtrsim$ 2550\,K) planets MASCARA-1b, TOI-1518b, HAT-P-70b, WASP-189b, and KELT-9b~\citep{bello-arufe_mining_2022, scandariato_pepsi_2023, prinoth_titanium_2022, borsato_mantis_2023, guo_detection_2024, simonnin_time-resolved_2025}. 
Yet another is SiO being in the gas phase on T$_{\mathrm{eq}}$ $\gtrsim$ 2200\,K planets but not on HAT-P-41b (T$_{\mathrm{eq}}$ = 1950\,K), likely due to condensation~\citep{lothringer_uv_2022, chachan_strong_2025}.

\begin{table*}
\caption{Overview of the NIRPS observations of WASP-189b.  Values in parentheses correspond to the simultaneously obtained HARPS exposures.  \label{tab:observations}}
\vspace{-6mm}
\centering
\begin{tabular}{ccccccc}
\hline
\hline
Date  &  Duration [h]  &  \# of exposures   &  $T_{\mathrm{exp}}$ [s]   &  Mean S/N  & Orbital phase range & Airmass\\
\hline
2023-06-03 & 6.1 & 53 (51) & 400 (400) & 120 (80) &  0.60 -- 0.69 & $1.58 \rightarrow 1.11 \rightarrow 1.54$  \\
2023-06-05 & 6.0 & 102 (93) & 200 (200) & 80 (60) &  0.33 -- 0.42 & $1.59 \rightarrow 1.11 \rightarrow 1.51$  \\
\hline
\hline
\end{tabular}
\vspace{-1mm}
\end{table*}

Characterising the sequential removal of metals is not only critical for accurately inferring the refractory elemental abundance of planetary envelopes in the context of planet formation, it also provides a unique opportunity to study how cold traps shape exoplanetary atmospheres. In particular, the degree of depletion of different metals with varying refractoriness (e.g.\ abundance of Ti relative to Fe or Fe relative to C) can not only serve as a probe of the composition of nightside clouds but also provide insights into the gravitational settling, advection, and vertical mixing timescales~\citep{parmentier_transitions_2016, helling_sparkling_2019, helling_cloud_2021}. For example, on the ultra-hot Jupiter WASP-76b (T$_{\mathrm{eq}}$ $\sim$ 2200\,K), elements such as V, Ba, and Ca have near-solar abundances compared to Ti which is depleted despite these elements having condensation temperatures within 150\,K~\citep{lodders_solar_2003}. This would suggest that a cold trap, once active, is highly efficient at removing a particular species from the observable gas phase~\citep{pelletier_vanadium_2023}. In contrast, the confirmed presence~\citep{prinoth_titanium_2025} but significant underabundance of Ti on the ultra-hot Jupiter WASP-121b (T$_{\mathrm{eq}}$ $\sim$ 2350\,K) suggests a cold-trap mechanism that is only partially efficient~\citep{pelletier_enriched_2026}. This would point towards the breaking of cold traps and the ensuing onset of metals in hot giant exoplanets being a progressive transition rather than a sharp one. 

The ultra-hot Jupiter WASP-189b~\citep[T$_{\mathrm{eq}}$ $\sim$ 2641\,K,][]{anderson_wasp-189b_2018} has had Ti, Ti$^+$, and TiO detected in its atmosphere~\citep{prinoth_titanium_2022, prinoth_time-resolved_2023}, suggesting that Ti-species are fully in the gas phase. However, with a poor day-to-night heat redistribution efficiency~\citep{deline_atmosphere_2022}, it may still be possible for Ti to condense and partially rainout in parts of the colder nightside. Previous works have mostly measured titanium to be slightly underabundant compared to model predictions on WASP-189b~\citep{gandhi_retrieval_2023, sanchez_stellar_2026} and other planets with similar equilibrium temperatures, such as MASCARA-1b~\citep{guo_detection_2024} and HAT-P-70b~\citep{gandhi_retrieval_2023, guo_elemental_2026}. Here we aim to independently measure the relative abundances of Fe and Ti on WASP-189b and interpret these in the context of cold trapping within the ultra-hot Jupiter population.

\section{Observations and methods} \label{sec:methods}

We observed the dayside of the ultra-hot Jupiter WASP-189b simultaneously with the HARPS~\citep{pepe_harps_2000} and NIRPS~\citep{bouchy_nirps_2025} spectrographs on the ESO 3.6m telescope in La Silla, Chile. The data were obtained after and before the secondary eclipse, on 3 June 2023 and 5 June 2025, respectively. Both time series lasted $\sim$6 hours, with details of the observations noted in Table~\ref{tab:observations}. The HARPS spectrograph covers the 0.38 -- 0.69\,$\mu$m wavelength region at a spectral resolution of $R \sim 110\,000$, while the NIRPS spectrograph simultaneously spans the $Y$, $J$, and $H$ bands (0.98 -- 1.8\,$\mu$m) at $R \sim 80\,000$. Observations were taken as part of the atmosphere characterisation portion~\citep{allart_nirps_2025} of the NIRPS GTO, with HARPS in high accuracy (HAM) and NIRPS in high efficiency (HE) mode.

All exposures were extracted using the standard Data Reduction Software (DRS) pipeline (version 3.0.0 for HARPS and 3.0.1 for NIRPS) adapted from ESPRESSO~\citep{pepe_espresso_2021, bouchy_nirps_2025}.  For our analysis, we used the produced order-by-order spectra without using the DRS-provided blaze or telluric corrections.  Spectra for both instruments are provided by default in the barycentric frame by the DRS, i.e.\ Doppler shifted to account for the motion of the telescope due to Earth's orbital motion and rotation relative to the WASP-189 system at the time of each exposure. While our analysis focuses on these data products, we also verified reducing the NIRPS data with the \texttt{APERO} pipeline~\citep{cook_apero_2022}, which produces similar results.

Our HARPS and NIRPS data build upon a wealth of previous observations of the atmosphere of WASP-189b in photometry~\citep{lendl_hot_2020, deline_atmosphere_2022, patel_tess_2026}, transmission~\citep{prinoth_titanium_2022, prinoth_time-resolved_2023, prinoth_atlas_2024, sreejith_cute_2023, vaulato_hydride_2025, borsato_untapped_2025}, and dayside thermal emission~\citep{yan_temperature_2020, yan_detection_2022b, deibert_high-resolution_2024, lesjak_retrieving_2025, van_sluijs_first_2025, sanchez_stellar_2026}.  Relative to these, our combined optical plus near-infrared wavelength coverage of the planetary dayside are well suited for simultaneous probing both the atomic and molecular budget of chemical species on the dayside, deeper in the atmosphere than typically probed by transit spectroscopy.

\begin{figure*}[!ht]
\begin{center}
\includegraphics[width=\linewidth]{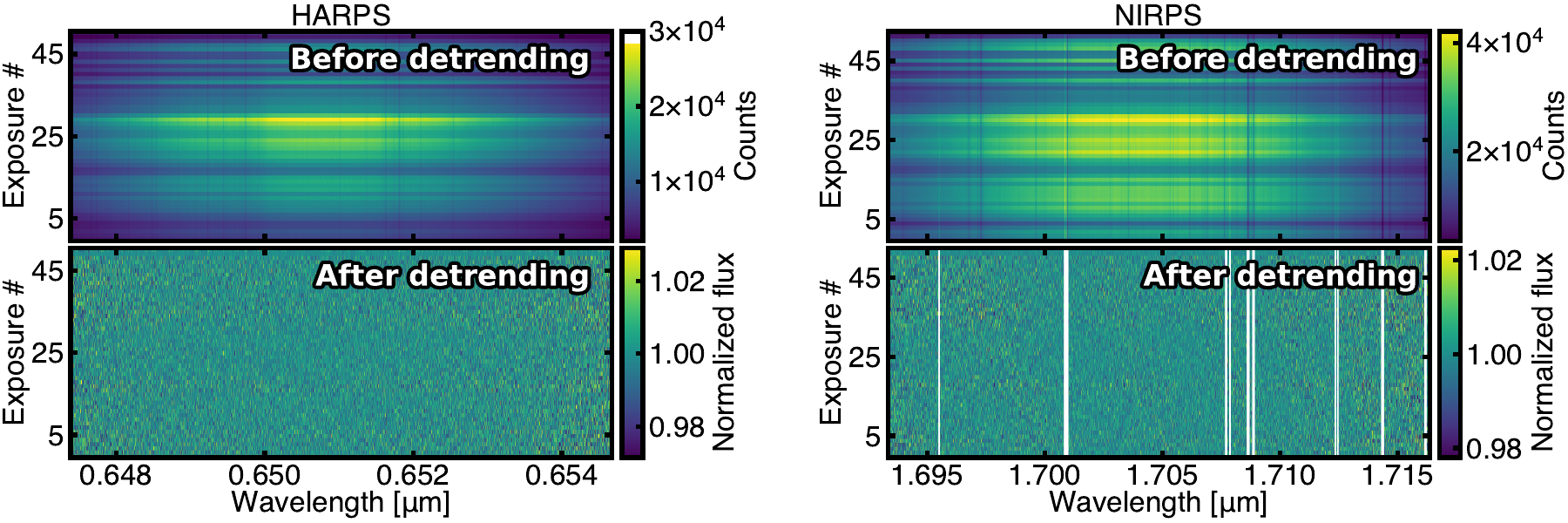}
\end{center}
\vspace{-5mm}
\caption{
Data before and after detrending for one example spectral order of HARPS (left) and NIRPS (right). The top panels show the dayside observations of WASP-189b on 3 June 2023 taken simultaneously with both instruments. Vertical stripes in the spectroscopic time series are telluric and stellar absorption lines (dark) or OH emission features (bright). Horizontal stripes are from variations in flux throughput as a result of changing observing conditions.  The bottom panels shows the data after removal of tellurics and stellar lines (see Section~\ref{sec:methods}). Vertical white stripes in the bottom-right panel mark wavelengths masked due to deep telluric absorption or OH emission lines. The absence of structured features in the data residuals demonstrates that the detrending (five principal components removed) effectively removes stellar and telluric lines. This `cleaned' data product, shown here for a single spectral order, was used for the cross-correlation and retrieval analyses.
}
\label{fig:data_reduction}
\end{figure*}

\subsection{Data detrending} \label{subsec:data_detrending}

To remove telluric and stellar contributions from the data and reveal the planetary signal of interest, the NIRPS and HARPS observations were processed as described in \cite{pelletier_crires_2025}. In brief, each time series data cube was corrected to remove bad pixels, aligned in either the stellar or telluric frame, masked to remove deep telluric absorption and emission lines, continuum aligned, median spectrum removed, and passed through a Principal Component Analysis (PCA). The HARPS data covering optical wavelengths were Doppler shifted to account for the reflex motion of the host star induced by the orbit of WASP-189b.  This is to ensure that stellar lines remain at fixed wavelengths for all exposures and thus are more efficiently removed by the median spectrum fit and PCA.  Conversely, the NIRPS data in the near-infrared were aligned in Earth's rest frame such that telluric line positions do not vary throughout the observations.  The reasoning for the different frame of reference alignment of the HARPS and NIRPS data is because stellar lines are the main source of contamination in the optical, while tellurics are comparatively more pronounced in the near-infrared. Detrending with PCA is optimal to remove spectral contributions that do not vary in wavelength (i.e.\ stellar lines in the stellar rest frame and telluric lines in the telluric rest frame) all the while leaving the rapidly accelerating planetary signal only partially affected in a manner that can be accounted for in the modelling~\citep{brogi_retrieving_2019}.

We set the threshold for telluric masking at a 50\% transmittance (i.e.\ all wavelengths where Earth's atmosphere blocks more than half the incoming light are masked), in addition to masking the entirety of the 1.345–1.446\,$\mu$m region as well as all wavelengths beyond 1.80\,$\mu$m where H$_2$O has strong absorption bands. We also mask telluric emission lines, mostly from OH and O$_2$, by discarding all wavelengths matching known features from sky observations, as tabulated by \cite{oliva_lines_2015}. For the PCA, we removed five principal components, although we tested removing between three and ten, finding similar results in all cases. For our analysis, all detrending steps are applied on each spectral order separately, treating each independently as its own time series.  Example orders of the HARPS and NIRPS data before and after detrending steps are shown in Figure~\ref{fig:data_reduction}. We also estimate the time dependent uncertainty ($\sigma_i$) of every pixel by fitting a noise model as described in \cite{gibson_detection_2020, gibson_relative_2022}.

\subsection{Forward atmosphere model} \label{subsec:fwd_model}

We used the SCARLET framework~\citep{benneke_atmospheric_2012, benneke_how_2013,  benneke_strict_2015} to generate synthetic models of the dayside atmosphere of WASP-189b. We first computed a baseline model for the cross-correlation analysis, which assumes a temperature structure in self-consistent radiative-convective equilibrium and a solar-like composition in chemical equilibrium with abundance profiles computed using  \texttt{FASTCHEM\,COND}~\citep{kitzmann_fastchem_2024}. The line opacity contributions considered in this work are Fe, Fe$^+$, Ti, Ti$^+$~\citep{kurucz_including_2017}, TiO~\citep{mckemmish_exomol_2019, mckemmish_hybrid_2024}, TiH~\citep{burrows_spectroscopic_2005, bernath_mollist_2020}, and FeH~\citep{wende_crires_2010}. 
Cross-sections were computed with \texttt{HELIOS-K}~\citep{grimm_helios-k_2015, grimm_helios-k_2021} for the molecules and ions. For the neutral atomic metals, we use the cross-sections from the \texttt{petitRADTRANS}~\citep{molliere_petitradtrans_2019} opacity database\footnote{\url{https://petitradtrans.readthedocs.io/en/latest/content/available\_opacities.html}}. Continuum opacity contributions comprise H$^-$ bound-free and free-free~\citep{gray_observation_2021, vaulato_hydride_2025}, as well as collision induced absorption from H$_2$, He, and H interactions~\citep{gustafsson_infrared_2001, gustafsson_h2h_2003, abel_collision-induced_2011, abel_infrared_2012}. The H$^-$ bound-free opacity is set by the abundance of H$^-$, while the free-free opacity by the product of the electron pressure and the neutral atomic H abundance. 

The emission spectrum of WASP-189b calculated from an atmospheric grid of 50 layers uniformly distributed in $\log$ pressure between 10$^1$ and 10$^{-8}$\,bar is then generated at a spectral resolution of $R = \lambda / \Delta \lambda  =$ 250\,000 and convolved with a rotational broadening kernel computed using Eq.~2 from \cite{reiners_feasibility_2002} assuming a tidally locked equatorial velocity of $V_{\mathrm{rot}} = 2\pi R_p / P = 2.99$\,km\,s$^{-1}$ and no limb darkening ($\epsilon = 0$). Our choice to not enforce any limb darkening is because, unlike for stars and colder exoplanets, ultra-hot Jupiters can also show limb brightening if they have thermally inverted atmospheres.

\subsection{Cross-correlation} \label{subsec:ccf_method}

A cross-correlation analysis is our method of choice to initially explore and visually identify the presence of individual chemical species in the dayside atmosphere of WASP-189b. Rather than using a stellar template as in the retrievals, we compute the planet-to-star flux ratio assuming the host star spectrum to be a blackbody for the cross-correlation. This is done to ensure that any identified signal is due to correlation with planetary rather than stellar lines while still approximately replicating the expected relative contrast of lines as a function of wavelength. We then remove the model continuum by dividing out an atmosphere model generated without any line opacities and convolve it with a Gaussian kernel matching the instrumental resolutions of either HARPS ($R$ = 110\,000) or NIRPS ($R$ = 80\,000).

With a model in hand, the cross-correlation function (CCF),
\begin{equation} \label{eq:ccf}
  \mathrm{CCF}(v) = \sum_i \frac{d_i m_{i}(v)}{\sigma_i},
\end{equation}
between cleaned data residuals ($d$) and model ($m$) at every pixel ($i$) is computed over a range of velocity shifts between $-400$ and $400$\,km\,s$^{-1}$ in steps of $1$\,km\,s$^{-1}$ ($v$) for each observed exposure~\citep{gibson_detection_2020}. The CCFs of both visits are then shifted to the rest frame of WASP-189b for the expected Keplerian velocity (Table~\ref{tab:system_params}), interpolated to a common orbital phase grid, and then summed. In the presence of an atmospheric signal, the combined phase-resolved CCF in the planetary rest frame will show the orbital trace of the planet closely aligned with the expected Keplerian motion of WASP-189b at the time of the observations (e.g.\ Figure~\ref{fig:CCF_trail}).  Deviations from the known orbit can then be interpreted as the result of uncertainties in the system parameters (e.g.\ $K_p$, $V_{\mathrm{sys}}$, mid-eclipse time), three dimensional effects, or atmospheric dynamics~\citep[e.g.][]{beltz_magnetic_2022, wardenier_pretransit_2025}. This cross-correlation process is repeated for single-species templates, i.e.\ models generated with only one individual, non-continuum opacity source at a time.

Any CCF signal can also be stacked in time via phase folding to produce $K_p$-$V_{\mathrm{sys}}$ maps, which can help identify fainter signals and provide an estimate of the signal-to-noise ratio (S/N) of a detection. Here the S/N is determined by dividing a cross-correlation map by its standard deviation excluding velocities within 30\,km\,s$^{-1}$ of the expected signal location. We note that the exact S/N is dependent on the somewhat arbitrarily chosen extent of the generated $K_p$-$V_{\mathrm{sys}}$ map and size of the excluded region.  The S/N scaling should therefore not be interpreted as a statistically robust estimate of the detection significance, but rather a simple metric for identifying signals of interest significantly exceeding that of any random noise feature.

\subsection{Retrieval prescriptions} \label{subsec:retrieval_method}

While a cross-correlation approach is suitable for identifying atmospheric absorbers, we used a Bayesian retrieval approach to quantitatively characterise the composition and temperature structure of the dayside of WASP-189b. For the atmospheric composition, we make use of two different retrieval flavours: `free' (or well-mixed), and chemical equilibrium. 

The free retrieval fits for each species as its own parameter assuming constant-with-altitude abundance profiles, with the exception of TiO and H$^{-}$ which are parameterised following \cite{parmentier_thermal_2018} to allow for decreasing abundances at high temperatures and low pressures due to thermal dissociation. Specifically, we fit for the $\log_{10}$ abundances of Fe, Fe$^+$, Ti, TiO, Ti$^+$, H$^-$, and $e^-$. To ensure that the volume mixing ratio of all species sum to unity in all atmospheric layers, we use a combination of H$_2$, He, and H as filler gases in proportions calculated by \texttt{FASTCHEM\,COND} for each queried model. A free retrieval has the advantage of being more data-driven, able to explore any chemical composition, with the downside of not accounting for some expected physical effects such as ionisation. 

The chemical equilibrium retrieval fits for the $\log_{10}$ elemental enrichment of Fe and Ti relative to solar ($[$Fe/H$]_{\odot}$ and $[$Ti/H$]_{\odot}$). While abundance profiles are predicted from a full chemical network of species, we only include opacities for Fe, FeH and Fe$^+$ (controlled by $[$Fe/H$]_{\odot}$) and Ti, TiO, TiH, and Ti$^+$ (controlled by $[$Ti/H$]_{\odot}$) mainly for computational reasons. Here a value of $[$Fe/H$]_{\odot}$ = 0 refers to an Fe abundance equal to the \cite{asplund_chemical_2009} solar value which is the reference in \texttt{FASTCHEM\,COND}. A chemical equilibrium approach has the benefit of coupling abundances of given species both together and to the temperature profile, while also naturally accounting for non-uniform abundance profiles due to the ionisation of metals and thermal dissociation of molecules without introducing numerous free parameters. However, it has the disadvantage of being constrained to only exploring atmospheric compositions allowed by equilibrium chemistry.

For both retrieval prescriptions, we fit for the vertical thermal structure with eight points uniformly distributed in $\log$ pressure between 10$^1$ and 10$^{-6}$\,bar. The temperature at each point can freely vary, with a smoothing prior on the second derivative with respect to the $\log_{10}$ pressure of 300\,K\,dex$^{-2}$ using the parameterisation of \cite{pelletier_where_2021} to prevent unphysical lapse rates.  Due to poor constraints at low pressures, we only fit the temperature profile until 10$^{-6}$\,bar, with the temperature assumed to be isothermal down to a pressure of 10$^{-8}$\,bar. The velocity parameters $V_{\mathrm{sys}}$ and $K_p$ are also kept as free parameters in the retrievals. Uniform priors are assumed for all parameters (Table~\ref{tab:retrieval_results}).

\begin{table}
\caption{Star and planet properties.} 
\vspace{-3mm}
\label{tab:system_params}
\centering
\def\arraystretch{1.1}
\begin{tabular}{ccc}
\hline
\hline
Parameter & Value & Reference \\
\hline
\hline
$R_*$ (R$_{\mathrm{\odot}}$) & $2.365 \pm 0.025$ & \cite{deline_atmosphere_2022} \\
$T_\mathrm{eff}$ (K) & $8000 \pm 80$ & \cite{lendl_hot_2020} \\
$\log g$ & $3.9 \pm 0.2$ & \cite{lendl_hot_2020} \\
$[$Fe$/$H$]$ & $0.29 \pm 0.13$ & \cite{lendl_hot_2020} \\
$v$sin$i$\,(km\,s$^{-1}$) & $93.1 \pm 1.7$ & \cite{lendl_hot_2020} \\
$K_{*}$ (km\,s$^{-1}$) & $0.182 \pm 0.013$ & \cite{yan_temperature_2020} \\
$V_{\mathrm{sys}}$ (km\,s$^{-1}$) & $-20.82 \pm 0.07$ & \cite{yan_temperature_2020} \\
\hline
$P$ (days) &  $2.7240308(28)$ & \cite{ivshina_tess_2022} \\
$T_0$\,(BJD$_\mathrm{TDB}$) & $2456706.4566(23)$ &  \cite{ivshina_tess_2022} \\
$R_p$ (R$_{\mathrm{Jup}}$) & $1.600_{-0.016}^{+0.017}$  & \cite{deline_atmosphere_2022}  \\
$M_p$ (M$_{\mathrm{Jup}}$) & $1.99_{-0.014}^{+0.016}$ & \cite{lendl_hot_2020}  \\
$K_p$ (km\,s$^{-1}$) & $200.7 \pm 4.9$  &  \cite{prinoth_titanium_2022} \\
$V_{\mathrm{rot}}$ (km\,s$^{-1}$) & $2.99 \pm 0.03$ & Assumed tidally locked  \\
\hline
\end{tabular}\\ 
\vspace{-3mm}
\end{table}

Unlike the cross-correlation which primarily aims to identify rather than quantitively characterise spectral features and could be done with a binary mask, atmospheric retrieval analyses require accurate modelling of the observed planetary spectrum. This thus necessitates using a more realistic stellar model, and mimicking on the model any manner in which the data detrending may have affected the underlying planet signal. For the stellar spectrum, rather than using a blackbody as for the cross-correlation analysis, we use a PHOENIX model~\citep{husser_new_2013} interpolated to match the T$_\mathrm{eff}$ and $\log g$ of the host star (Table~\ref{tab:system_params}) and rotationally broadened to a $v$sin$i$ of 93.1\,km\,s$^{-1}$ assuming a limb darkening coefficient $\epsilon = 1$. The broadened stellar model is then Doppler shifted to the expected velocity shift ($\Delta v$) needed to align spectral lines with associated feature positions in the data for every observed exposure. For the NIRPS data which is in the telluric rest frame, the shift accounts for $V_\mathrm{sys}$, the barycentric Earth radial velocity (BERV), and WASP-189's reflex motion due to the orbital motion of WASP-189b.  For the HARPS data sets already corrected for the BERV and stellar wobble, this involves a uniformly shifting every exposure by $V_\mathrm{sys}$. Conversely to assuming a blackbody, using a stellar template requires correctly accounting for the stellar line positions at each exposure time to accurately compute the observed wavelength dependent planet-to-star flux ratio at each pixel.  

For each queried $K_p$ and $V_{\mathrm{sys}}$ in the retrieval, the modelled planet spectrum $F_p$ is projected in phase for each observed time series and scaled to the stellar flux contrast ratio via
\begin{equation} \label{eq:fpfs}
    F_{\mathrm{scaled}} = \frac{F_p(K_p,V_{\mathrm{sys}})}{F_s(\Delta v)} \left( \frac{R_p}{R_s} \right)^2.
\end{equation}
The scaled model is then convolved first with a rotational kernel assuming a tidally locked equatorial velocity of 2.99\,km\,s$^{-1}$, and then with a Gaussian broadening kernel matching the instrumental resolutions of HARPS ($R$ = 110\,000) or NIRPS ($R$ = 80\,000).

As the underlying planetary signal may be stretched and scaled during the data detrending procedure using PCA, the effect must be reproduced on the models to ensure accurate parameter inferences~\citep{brogi_retrieving_2019}. For this, we use the fast model-filtering technique of \cite{gibson_relative_2022}, which is numerically almost identical but more computationally efficient than re-applying PCA on every modelled time series. With the filtered model, we then computec the likelihood $\mathcal{L}$ based on the formulation from \cite{gibson_detection_2020} with the maximum likelihood optimised noise scaling term,
\begin{equation} \label{eq:likelihood}
         \ln\mathcal{L} = -\frac{N}{2} \ln\left[\frac{1}{N}\left(\sum_{i} \frac{d_{i}^2+ m_{i}^2}{\sigma^{2}_{i}}-2\mathrm{CCF}\right)\right],
\end{equation}
where $N$ is the number of pixels and the model is assumed to be correctly scaled. Note that $m_i$ here is the processed (filtered) model, in contrast to Eq.~\ref{eq:ccf} where we use a more agnostic un-processed model for the cross-correlation analysis. To sample the parameter space in the retrievals, we use the Markov chain Monte Carlo package \texttt{emcee}~\citep{foreman-mackey_emcee_2013}.

\section{Results and discussion}\label{sec:results}

\begin{figure}
\begin{center}
\includegraphics[width=\linewidth]{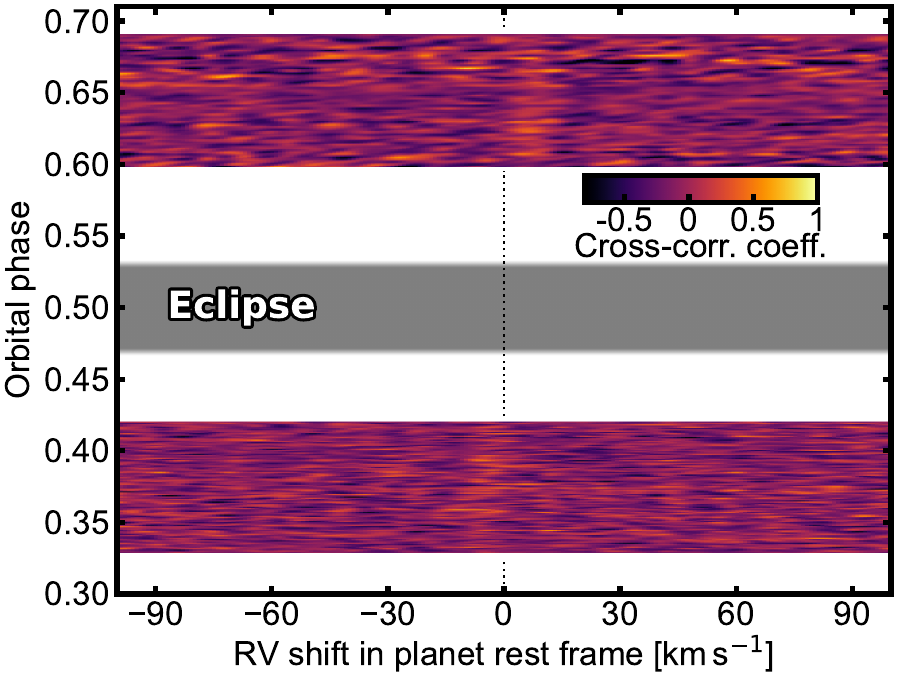}
\end{center}
\vspace{-5mm}
\caption{
Rest-frame orbital trace of WASP-189b as a function of phase observed from HARPS and NIRPS. The normalized cross-correlation map (coloured) is shifted in the planetary rest frame, assuming $K_p = 200.7$\,km\,s$^{-1}$ and $V_\mathrm{sys} = 20.82$\,km\,s$^{-1}$ such that the signal from a static atmosphere on a non-rotating planet should be at 0\,km\,s$^{-1}$ (dotted black line).
The observed signal of WASP-189b (vertical yellow streak) appears sloped, slightly blueshifted during pre-eclipse phases, and slightly redshifted after the eclipse.  This is likely a combination of the planetary rotation, eastward equatorial winds, and the true $K_p$ of WASP-189b being slightly lower than assumed.
}
\label{fig:CCF_trail}
\end{figure}

We cross-correlate the detrended HARPS and NIRPS data with a model template to search for a signature matching the expected orbital motion of WASP-189b. Viewed in the planetary rest frame, we detect a trace in both the pre- and post-eclipse visits closely matching the expected orbital motion of WASP-189b (Figure~\ref{fig:CCF_trail}).  The CCF gives a positive correlation when using a model with emission lines, indicating a thermally inverted atmosphere, as expected for ultra-hot Jupiters and consistent with previous works on WASP-189b~\citep{yan_temperature_2020, yan_detection_2022b, deibert_high-resolution_2024, lesjak_retrieving_2025, van_sluijs_first_2025}. 

\begin{figure*}
\begin{center}
\includegraphics[width=\linewidth]{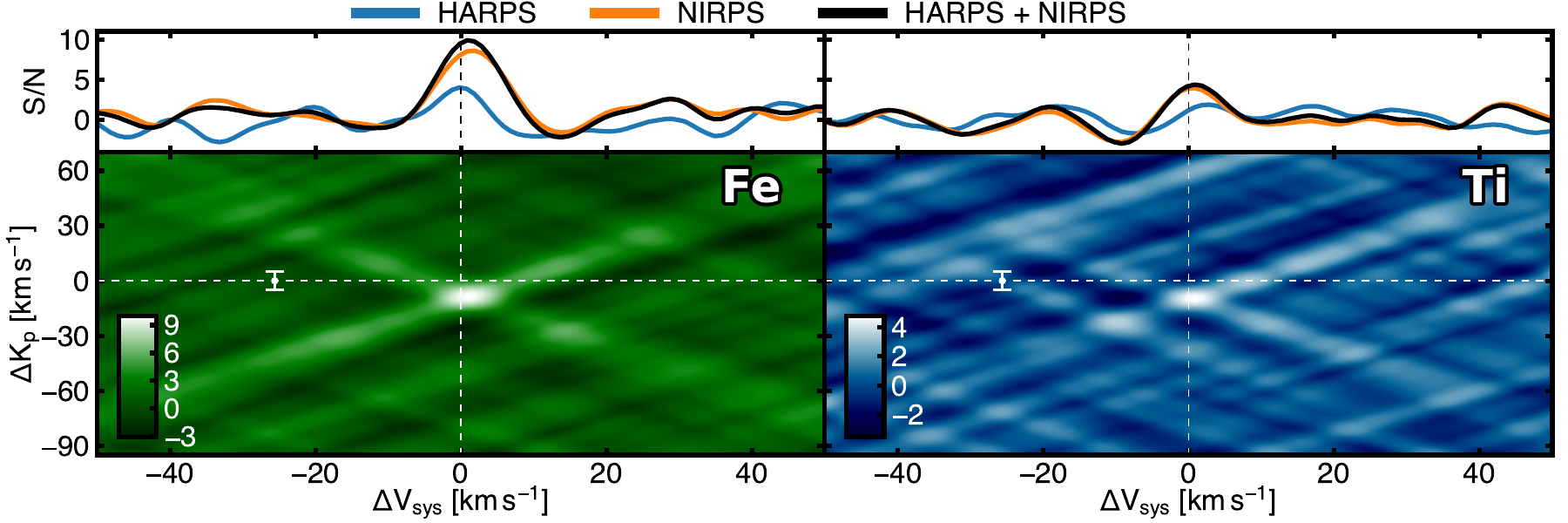}
\end{center}
\vspace{-5mm}
\caption{
Observed cross-correlation signals of Fe and Ti on the dayside atmosphere of WASP-189b.
Bottom panels: Two-dimensional HARPS $+$ NIRPS cross-correlation signal-to-noise map for Fe (left) and Ti (right) relative to deviations from the expected $K_p = 200.7 \pm 4.9$\,km\,s$^{-1}$ and $V_{\mathrm{sys}} = -20.82 \pm 0.07$\,km\,s$^{-1}$ (dashed white lines). While a negative $\Delta K_p$ value is predicted from 3D models, it remains consistent with the expected $K_p$ due to its relatively large measurement uncertainty (white error bar).
Top panels: CCF slices along the $V_{\mathrm{sys}}$ axis at $K_p$ = 193\,km\,s$^{-1}$ showing the contributions from HARPS (blue) and NIRPS (orange).  Despite the opacity from metals being stronger in the optical, most of the signal is driven by NIRPS owing to the better planet-to-star flux contrast at longer wavelengths.
}
\label{fig:CCF_maps}
\end{figure*}

Cross-correlating with atmospheric templates of different Fe- and Ti-bearing species, we detect a clear signal of neutral atomic Fe and a tentative signal of neutral atomic Ti near the expected $K_p$ and $V_\mathrm{sys}$ of WASP-189b (Figure~\ref{fig:CCF_maps}). The observed signals are primarily driven by the NIRPS data, although HARPS still contributes positively to the overall signals.  These thermal emission signals being stronger in near-infrared rather than optical data is in contrast to transit studies of ultra-hot Jupiters, which tend to show very strong signals for metals and ions in the optical~\citep[e.g.][]{merritt_inventory_2021, kesseli_atomic_2022, prinoth_atlas_2024} but more muted near infrared spectra~\citep{vaulato_hydride_2025}. Indeed, atomic and ionic species tend to have most of their spectral transitions at shorter wavelengths and transmission spectroscopy depends on the change in planet-to-star area ratio as a function of wavelength.  This makes infrared wavelengths that have fewer strong lines generally less favourable for detecting metals in transit studies. For emission spectroscopy, however, a planet's observability depends more on the wavelength dependent planet-to-star flux (rather than area) ratio.  Therefore, although neutral metals and ions typically have fewer and weaker spectral features at near-infrared wavelengths than in the optical, the planet itself emits proportionally much more light compared to the star. For the WASP-189 system specifically, assuming a planet dayside temperature of 3435\,K~\citep{lendl_hot_2020}, the planet-to-star contrast ratio (Eq.~\ref{eq:fpfs}) is approximately 18 times more favourable at 1.5\,$\mu$m than at 0.5\,$\mu$m.  Compared to transmission spectroscopy, the much better flux contrast of thermal emission observations at longer wavelengths can therefore compensate for the generally weaker spectral features of neutral atomic metals such as Fe and Ti in terms of detectability.

Other than neutral atomic Fe and Ti, we do not observe a significant cross-correlation signal for any other Fe- or Ti-bearing species (FeH, Fe$^+$, TiO, TiH, or Ti$^+$, see Figure~\ref{fig:CCF_non_detections})).  However, this does not mean that these species are not present on the dayside atmosphere of WASP-189b. Rather, it likely reflects the lack of sensitivity of our observations to these species. For example, Fe$^+$ and Ti$^+$ mostly have spectral features at the bluest wavelengths of the HARPS data where the thermal emission planet-to-star flux contrast is very low.  Indeed, we expect these species to be present on the dayside, given their previous detection in the transmission spectrum of WASP-189b~\citep{prinoth_titanium_2022, prinoth_time-resolved_2023}. While here we focus on Fe and Ti species, a full exploration of the spectral inventory of other chemical elements on the dayside atmosphere of WASP-189b in the near-infrared will be done in future work.

\subsection{Velocity offsets}\label{subsec:velocties}

The observed signature of WASP-189b in cross-correlation is measured at a $V_{\mathrm{sys}} = -19.9 \pm 1.8$\,km\,s$^{-1}$ and $K_p = 192.7 \pm 2.6$\,km\,s$^{-1}$, which is consistent with the stellar systemic velocity $V_\mathrm{sys}$ = $-20.82 \pm 0.07$\,km\,s$^{-1}$ measured by \cite{yan_temperature_2020}, but lower than the planetary orbital velocity semi-amplitude $K_p$ = $200.7 \pm 4.9$\,km\,s$^{-1}$~\citep{prinoth_titanium_2022}.  The position of the Fe and Ti signatures are consistent with previous high-resolution detections of the dayside atmosphere of WASP-189b~\citep{yan_temperature_2020, yan_detection_2022b, deibert_high-resolution_2024, van_sluijs_first_2025, lesjak_retrieving_2025, sanchez_stellar_2026}.

While blueshifts measured in transit observations that probe terminator regions can be indicative of day-to-night winds, the velocity component of an equatorial jet and day-to-night winds will be largely perpendicular to the line of sight for thermal emission observations of the integrated dayside of the planet when near secondary eclipse.  As a result, any observed net blue or redshift will not necessarily directly relate to wind speeds, and should show an orbital phase dependence as different parts of the atmosphere rotating towards or away from the observer come into view. For typical hot Jupiters, the Doppler imprint of an equatorial jet and an offset hotspot on thermal emission observations is expected to be relatively small ($\sim$3\,km\,s$^{-1}$ at most at post-eclipse phases) based on GCM models~\citep[][see their Figure 7]{zhang_constraining_2017}. Although this effect could potentially be larger for planets under more extreme irradiation conditions, GCM simulations of ultra-hot Jupiters generally do not predict significant $V_\mathrm{sys}$ offsets for observations near secondary eclipse~\citep{beltz_magnetic_2022, lee_3d_2022, wardenier_pretransit_2025}. Indeed, dayside thermal emission observations of ultra-hot Jupiters at high spectral resolution tend to show signals that are well aligned with the stellar systemic velocity~\citep[e.g.][]{cont_atmospheric_2022, brogi_roasting_2023, smith_roasting_2024, guilluy_gaps_2025, ramkumar_new_2025}. Nevertheless, the lack of an offset in $V_\mathrm{sys}$ between planet and star observed here in the dayside geometry should not be interpreted as the lack of winds on WASP-189b.  For example, transit observations of WASP-189b probing the terminator have measured significant winds~\citep{prinoth_titanium_2022, prinoth_time-resolved_2023, gandhi_retrieval_2023, vaulato_hydride_2025} while photometric phase curve observations also show some evidence of an equatorial jet-driven eastward-shifted hotspot~\citep{deline_atmosphere_2022, patel_tess_2026}.

Although planets move along their orbits at a set velocity, atmospheric signal observed from dayside observations may differ from the line-of-sight velocity semi-amplitude ($K_p$) inferred from radial velocity measurements.  Owing to the rotation of the planet, the substellar point and other parts of the dayside will rotate towards or away from the observer. This can cause deviations between the observed and true $K_p$, with the former appearing lower due to the planet rotation counteracting the orbital motion at pre-and post-eclipse phases~\citep[][see their Figure 8]{hoeijmakers_mantis_2024}. The magnitude of $\Delta K_p$ incurred as a result of rotation is expected to be of the order of the equatorial velocity, although this can be further amplified by an equatorial jet that acts in the same direction as the planetary rotation~\citep{wardenier_pretransit_2025, zhang_extreme_2026}. For WASP-189b, the cross-correlation signal of the dayside atmosphere ($K_p = 192.7 \pm 2.6$\,km\,s$^{-1}$) is lower than the expected value of $K_p$ = $200.7 \pm 4.9$\,km\,s$^{-1}$ by more than the tidally locked equatorial velocity ($V_{\mathrm{rot}}\sim$3\,km\,s$^{-1}$). The low $K_p$ could be the results of strong eastwards (i.e.\ in the direction of the planetary rotation) equatorial winds, or that the true $K_p$ is smaller, which could have implications for atmospheric studies via the assumed planetary mass. However, the $K_p$ values are still consistent within 1.5$\sigma$ considering the relatively large uncertainties of the measurements. We note that our results are in better agreement with the calculated $K_p = 197.8 \pm 5.8$\,km\,s$^{-1}$ from the parameters of \cite{anderson_wasp-189b_2018}.

Inferring dynamical and 3D effects from velocity offsets in high resolution observations relies on a comparison to the true $V_\mathrm{sys}$ and $K_p$. For rapidly rotating stars that have blurred spectral features such as WASP-189, however, obtaining precise and accurate estimates from radial velocity measurements is particularly challenging~\citep[e.g.][see their Section 3.1.1]{pai_asnodkar_kelt-9_2022}. Furthermore, reported uncertainties on $V_\mathrm{sys}$ measurements of fast rotators in the literature tend to be small, likely significantly underestimated. For example, \cite{anderson_wasp-189b_2018} found a $V_\mathrm{sys}$ = $-24.452 \pm 0.012$\,km\,s$^{-1}$, while \cite{yan_temperature_2020} reported $V_\mathrm{sys}$ = $-20.82 \pm 0.07$\,km\,s$^{-1}$ with both having used HARPS-N data.  These measurements taken with the same instrument being inconsistent at $\sim$50$\sigma$ suggests that their uncertainties are underestimated. For this work, we compare our planetary signal position to the $V_\mathrm{sys}$ measured from \cite{yan_temperature_2020} as it is more consistent with the measured $V_\mathrm{sys}$ values from ESPRESSO and MAROON-X observations~\citep{seidel_magnetic_2026}.  Alternatively, using $V_\mathrm{sys} = -24.45$\,km\,s$^{-1}$ would imply a planetary signal redshifted by $\sim$4\,km\,s$^{-1}$, which is more difficult to explain for dayside thermal emission observations.

\begin{figure*}
\begin{center}
\includegraphics[width=\linewidth]{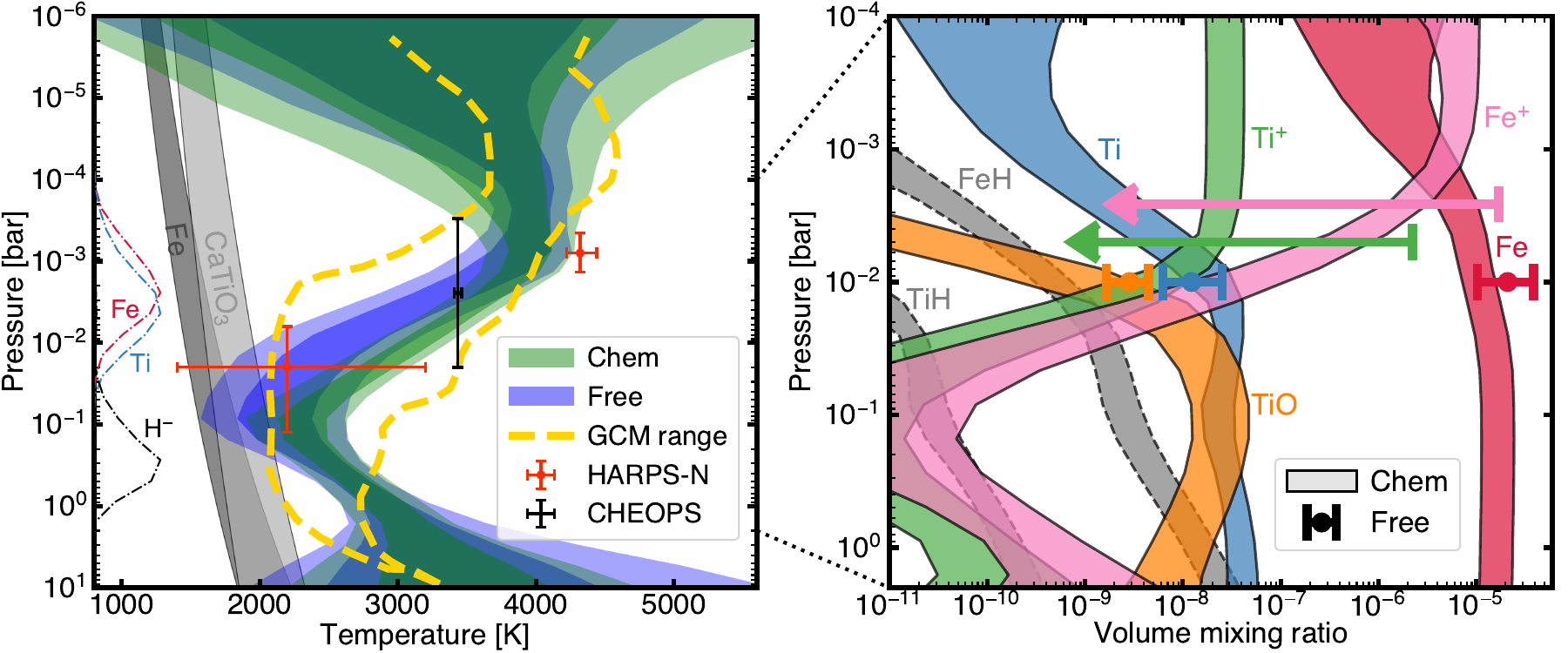}
\end{center}
\vspace{-5mm}
\caption{
Retrieved dayside vertical temperature structure and atmospheric composition of WASP-189b determined from HARPS and NIRPS.  Left panel: Recovered thermal profile (1 and 2-$\sigma$ contours) from the free (blue) and chemical equilibrium (green) retrievals compared to the range of dayside vertical TP profiles within $\pm$90$^\circ$ longitude of the substellar point from the GCM of \cite{lee_mantis_2022}. The two-point profile from HARPS-N~\citep{yan_temperature_2020} and the dayside temperature of 3542$\pm$14\,K measured from CHEOPS~\citep{lendl_hot_2020, deline_atmosphere_2022} are also shown for comparison, with uncertainties in pressure for the latter approximated to cover the extent of the GCM models. The light and dark grey contours respectively represent the TiO$_2$ and Fe condensation curves for atmospheric compositions between 0.1 and 10$\times$ solar~\citep{burrows_chemical_1999, visscher_atmospheric_2010}. Contribution functions showing the pressures probed by the primary opacity contributors are shown as dot-dashed lines. While the dayside of WASP-189b is too hot for any significant condensation to occur, some TiO$_2$ may still be able to condense and rainout on the colder night-side. Right panel: Inferred compositions from the free (data points) and chemical equilibrium (contours) retrievals. Note that the trace molecules FeH and TiH (dashed grey contours) are considered in the chemical equilibrium but not in the free retrieval.  Both prescriptions yield consistent constraints for Fe and Ti species at the average pressures probed ($\sim$1 -- 10\,mbar). 
}
\label{fig:TP_abun}
\end{figure*}

Beyond using an accurate stellar $V_\mathrm{sys}$, 
inferring any offsets of the planetary signal also relies on correctly knowing the planet's true velocity at the time of each observed exposure, to then compare to the data. For example, using an out-of-date ephemeris for the planetary orbit can result in even small uncertainties in the orbital period being propagated over many orbits, leading to an inaccurate assumed transit mid-point timing. Other factors that can lead to km\,s$^{-1}$ offsets in the $V_\mathrm{sys}$ of observed planetary signals include using timings that are not in Barycentric Dynamical Time (BJD$_{\mathrm{TDB}}$) and using the start rather than the mid time for exposures (Figure~\ref{fig:CCF_offsets}). Offsets in $K_p$ can also be incurred from neglecting even a small eccentricity (Appendix~\ref{appendix:KpVsys_tests}).

\subsection{Retrieval results}\label{subsec:retrieval_results}

We quantitatively inferred the composition, vertical temperature-pressure (TP) profile, and velocity parameters of the dayside atmosphere of WASP-189b using the free and chemical equilibrium retrieval setups described in Section~\ref{subsec:retrieval_method} (Table~\ref{tab:retrieval_results}). In both cases we recovered a $\sim$2000\,K-amplitude thermal inversion (Figure~\ref{fig:TP_abun}, left panel), consistent with GCM predictions~\citep{lee_3d_2022} and previous results~\citep{lendl_hot_2020, yan_temperature_2020, lesjak_retrieving_2025, sanchez_stellar_2026}.  

From the equilibrium chemistry retrieval, we obtained bounded constraints for both $[$Fe/H$]_{\odot}$ and $[$Ti/H$]_{\odot}$.  We measure these to be slightly sub-solar, with $[$Fe/H$]_{\odot}$ = $-0.44_{-0.16}^{+0.15}$, and $[$Ti/H$]_{\odot}$ = $-0.94_{-0.11}^{+0.12}$. Compared instead to the host star~\citep{lam_secrets_2024}, the planetary metal enrichments are significantly substellar, with $[$Fe/H$]_{*}$ = $-0.94_{-0.16}^{+0.15}$ and $[$Ti/H$]_{*}$ = $-1.26_{-0.11}^{+0.12}$. We note that inferred abundances in atmospheric retrievals can be strongly correlated to the temperature structure and the continuum, while abundance ratios should be more robust~\citep{benneke_atmospheric_2012}, especially in high-resolution studies~\citep{gibson_relative_2022}. For the relative proportion of iron and titanium, we find this to be slightly below that of the Sun ($[$Ti/Fe$]_{\odot}$ = $-0.51_{-0.22}^{+0.23}$) and marginally below the value of the host star of $[$Ti/Fe$]_{*}$ = $-0.33_{-0.23}^{+0.24}$~\citep{lam_secrets_2024}. Assuming that WASP-189b should have the Ti/Fe ratio of its host star, this would imply that only between 28\% and 81\% (1$\sigma$ bounds) of the titanium is accounted for in the observable dayside atmosphere.

Independently, the volume mixing ratios of individual chemical species inferred from the free retrieval are consistent with the abundance profiles predicted in the chemical equilibrium retrieval (Figure~\ref{fig:TP_abun}, right panel).  As expected, only weak constraints are placed on Fe$^+$ and Ti$^+$ (Figure~\ref{fig:corner_free}) due to the limited sensitivity of our data to these ions with spectral features mostly at short wavelengths. Interestingly, TiO is recovered from the free retrieval despite the lack of a clear signal seen in cross-correlation. This likely indicates that the constraints on the TiO abundance are driven more by its molecular bands shaping the overall spectrum rather than its line opacities~\citep[e.g.][]{vaulato_hydride_2025}.  
Indeed, while the chemical equilibrium retrieval predicts a relative proportion of Ti and TiO set by the temperature, the more flexible free retrieval interestingly also recovers similar abundances for both species. While our measured TiO abundance should be considered more of an upper limit given its non-detection in cross-correlation (Figure~\ref{fig:CCF_non_detections}), if we nevertheless assume that all Fe is in atomic form and all Ti is in either neutral or molecular form, we estimate a substellar $[$Ti/Fe$]_{*}$ = $-0.43_{-0.30}^{+0.31}$ from the free retrieval 
($[$Ti/Fe$]_{*}$ = $-0.54_{-0.33}^{+0.34}$ if only considering atomic Ti), slightly lower but overall consistent with the chemical equilibrium approach (Table~\ref{tab:retrieval_results}). The difference is likely due to ionic species not being included in the elemental abundance estimations of the free retrieval.  As Ti has a lower binding energy than Fe, more Ti will be ionised relative to Fe, which can bias the ratio to a smaller value.

\subsection{Potentially missing titanium}\label{subsec:titanium}

From a planet formation standpoint, rock-forming elements are expected to be condensed in all but the hottest innermost regions of the protoplanetary disc. As a result, refractory metals should always be accreted as solids and hence are expected to preserve their primordial relative abundances~\citep{lothringer_new_2021}. This is in contrast with volatile species such as O, C, and N, which may be accreted as gas or solids, depending on where in the disc a planet accretes its envelope~\citep{turrini_tracing_2021}. Therefore, while elemental abundance ratios such as C/Fe, C/O or O/N may vary relative to the stellar values depending on a planet's accretion history~\citep{chachan_breaking_2023}, this is not the case for refractory ratios (e.g.\ Ti/Fe). In the absence of observational biases, deviations between planetary and stellar refractory elemental ratios are likely due to atmospheric processes.  For example, nightside condensation and rainout can preferentially remove more refractory species from the gas phase, making them appear to be missing from the observable atmosphere~\citep{parmentier_transitions_2016, pelletier_vanadium_2023}.

\begin{figure}
\begin{center}
\includegraphics[width=\linewidth]{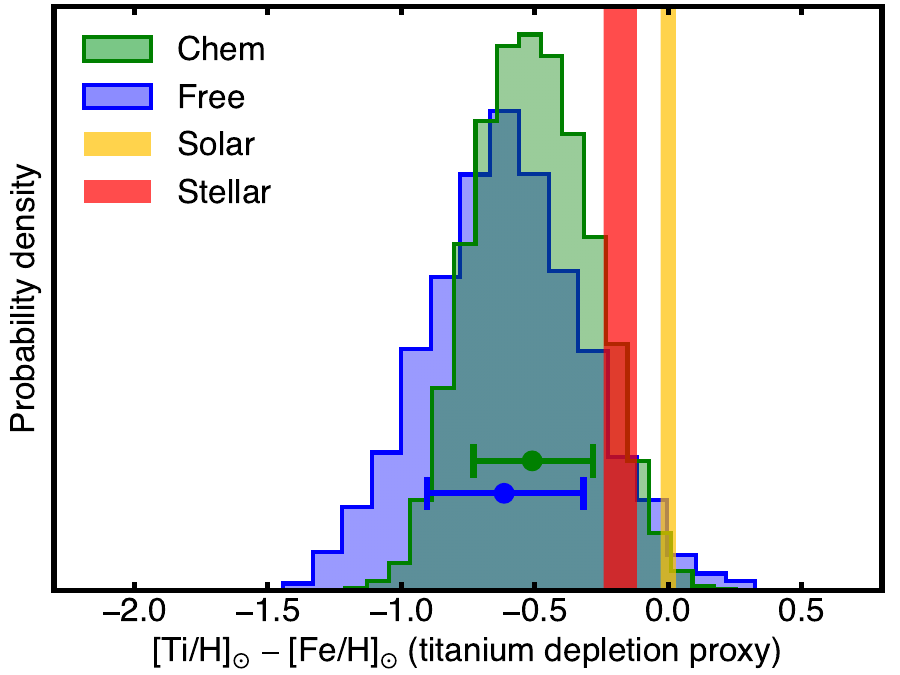}
\end{center}
\vspace{-5mm}
\caption{
Relative proportion of titanium and iron on WASP-189b from the free (blue) and chemical equilibrium (green) retrievals.  Stellar~\citep[][red]{lam_secrets_2024} and solar~\citep[][yellow]{asplund_chemical_2009} ratios are also shown for comparison. As Ti has a higher condensation temperature than Fe, the Ti/Fe ratio being slightly below the stellar and solar value could indicate that some titanium species are missing from the observable gas phase in the dayside atmosphere of WASP-189b. 
}
\label{fig:Ti_Fe}
\end{figure}

Working under this assumption, the substellar Ti/Fe ratio recovered by both our free and chemical equilibrium retrievals (Figure~\ref{fig:Ti_Fe}) would imply that not all of the titanium budget is accounted for on the dayside of WASP-189b. However, the underabundance of titanium relative to iron is only marginal, with our results still consistent with the stellar value at $\sim$2$\sigma$ and the solar value at $\sim$3$\sigma$.  While ionisation alone is unlikely to explain this slight observed underabundance, potential sources of bias in our measurements include the TiO line list being imperfect and the assumption of local thermodynamic equilibrium (LTE). In the case of the latter, however, our observations primarily probe high enough pressure regions of the atmosphere (>$10^{-4}$\,bar, Figure~\ref{fig:TP_abun}, left panel) that are not expected to be significantly affected by non-LTE effects~\citep{fossati_non-local_2021, fossati_gaps_2023}. With these caveats in mind, if we nevertheless assume that the substellar measured Ti/Fe is representative of the dayside atmosphere of WASP-189b, we can explore the possibility that part of the Ti-budget is missing due to nightside cold trapping.

\begin{figure*}
\begin{center}
\includegraphics[width=\linewidth]{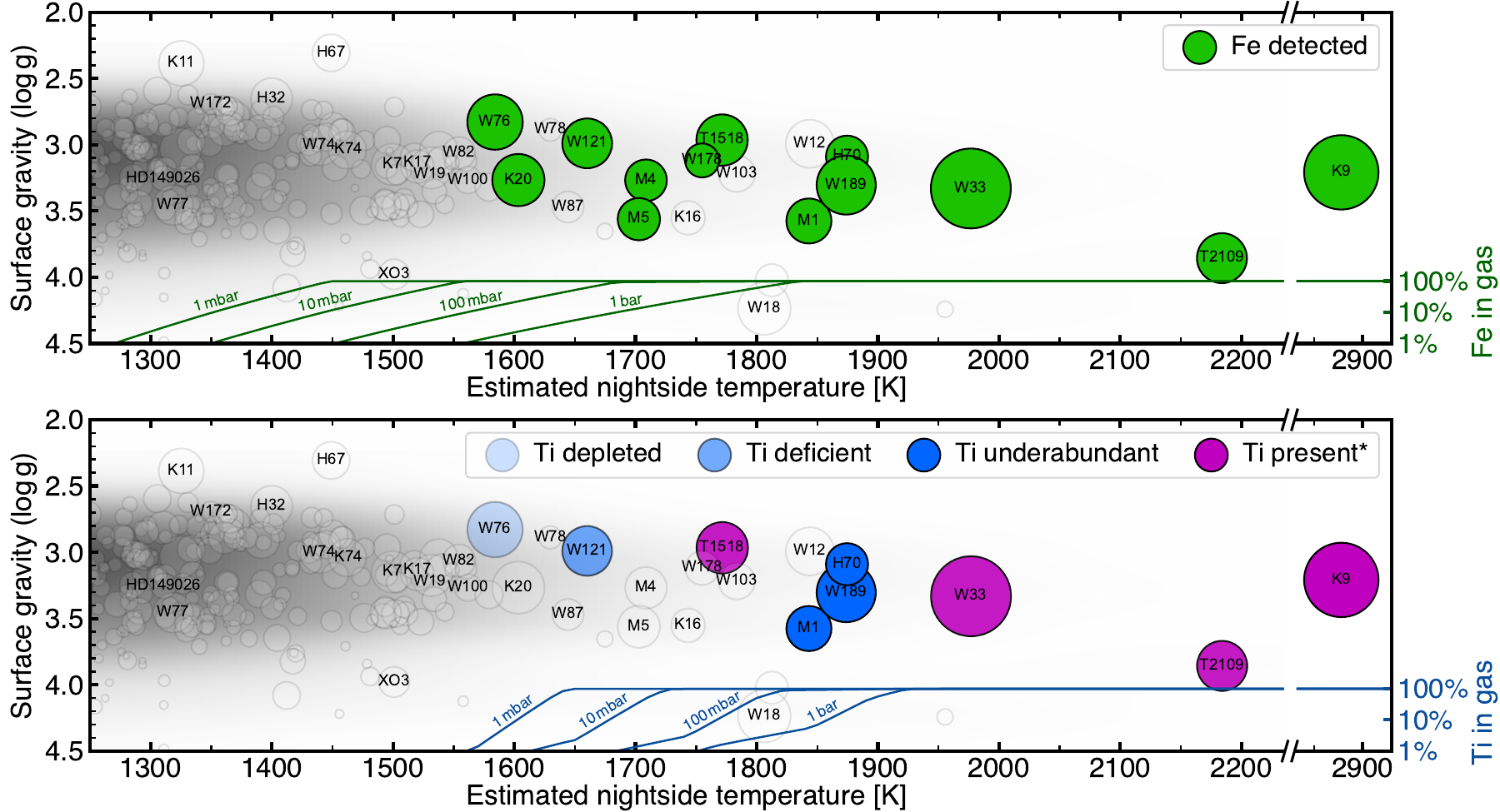}
\end{center}
\vspace{-5mm}
\caption{
Overview of iron and titanium detections in exoplanetary atmospheres compared to the overall exoplanet population (grey). Circle sizes correspond to an observational favourability estimate. Some exoplanet hosts are named for reference: `W'= WASP, `K' = KELT, `H' = HAT-P, `T' = TOI, and `M' = MASCARA. Top: Highlighted detections of Fe (green circles). The secondary axis shows the fraction of Fe-species (Fe, FeH, and Fe$^+$) in the gas phase predicted by \texttt{FASTCHEM\,COND} for a solar-composition gas at different pressure levels.  Iron is fully vapourised above $\sim$1800\,K and effectively condensed out of the gas phase below 1300\,K. The strong presence of Fe on WASP-76b would suggest that colder planets could also have gaseous Fe in their atmospheres. Bottom: Same but for Ti, with the secondary axis denoting the predicted gaseous fraction of Ti-species (Ti, TiO, TiO$_2$, and Ti$^+$).  The blue shading depicts the degree of depletion, while `Ti present$^*$' in purple refers to cases where Ti has been detected (e.g.\ via cross-correlation), but its abundance is not well characterised. Unlike for Fe, Ti shows evidence of significant depletion on WASP-76b, while its underabundance relative to predictions on the slightly hotter WASP-121b indicates that this planet may border the gaseous onset of titanium species in ultra-hot Jupiter atmospheres. Based on current detections the apparent onsets of both Fe and Ti roughly match the nightside temperature at which these species should vapourise at $\sim$10\,mbar, although this likely also depends on other parameters, such as the surface gravity, atmospheric metallicity, and heat redistribution.
}
\label{fig:population}
\end{figure*}

With a TP profile mostly above the condensation curve of even ultra-refractory TiO$_2$ (Figure~\ref{fig:TP_abun}, left panel), all forms of titanium on the dayside are expected to be fully vapourised.  However, even on a T$_\mathrm{eq} \sim$ 2650\,K tidally locked planet, it may be possible for clouds to form on the colder nightside, depending on the albedo and heat redistribution efficiency~\citep{helling_cloud_2021}.  As Ti is slightly more refractory than Fe and hence condenses more readily~\citep{lodders_solar_2003}, measuring their relative proportions on the dayside or terminator of a planet is a powerful diagnostic to indirectly probe for nightside cloud formation and rainout. A stellar-like Ti/Fe ratio would indicate a lack of cold trapping, while a depleted Ti/Fe ratio would point towards some Ti being removed from the gas phase.

\subsection{WASP-189b in context}\label{subsec:context}

To better understand the composition of WASP-189b in the context of the hot Jupiter population, we compile reported unambiguous detections of atomic Fe and Ti in the literature. We specifically do not include detections that are only tentative~\citep[e.g.\ Fe on WASP-172b,][]{seidel_detection_2023}, that have been contested~\citep[e.g.\ Fe and Ti on HD 149029b,][]{ishizuka_neutral_2021, biassoni_high-resolution_2024}, that may be due to Rossiter-McLaughlin residuals~\citep[e.g.\ Ti$^+$ on MASCARA-4b,][]{jiang_detection_2023}, or that are from TiO which are sometimes ambiguous and/or disputed~\citep[e.g.\ WASP-19b,][]{huitson_hst_2013, sedaghati_detection_2017, sedaghati_spectral_2021, espinoza_access_2019}. Based on current results, the onset of Fe in hot giant exoplanets is consistent with when it is expected to evaporate from its condensed state according to equilibrium chemistry calculations using \texttt{FASTCHEM\,COND}~\citep{kitzmann_fastchem_2024} based on the predicted nightside temperature (Figure~\ref{fig:population}, top panel).  Nightside temperatures are estimated from Eqs.\,4 \& 5 of \cite{cowan_statistics_2011} assuming the heat redistribution relation of \cite{parmentier_cloudy_2021} for the case of nightside clouds (their Eq.\,2 with $f$ clipped at a maximum value of 2.1, the approximate value of the hottest point in their grid). For reference, this estimates a nightside temperature of 1873\,K for WASP-189b, which is compatible with the T$_\mathrm{night} <$ 2100\,K limit from CHEOPS~\citep{deline_atmosphere_2022} and the T$_\mathrm{night}$ = $1529_{-209}^{+222}$\,K (with the authors noting that this is likely underestimated) inferred from TESS~\citep{patel_tess_2026}.

Compared to Fe, the observed onset of Ti in ultra-hot Jupiter atmospheres from depleted (Ti $\lesssim$ 1\%) on WASP-76b~\citep{pelletier_vanadium_2023}, to deficient (1\% $\lesssim$ Ti $\lesssim$ 10\%) on WASP-121b~\citep{pelletier_enriched_2026}, to slightly underabundant (10\% $\lesssim$ Ti $\lesssim$ 100\%) on T$_{\mathrm{eq}} \gtrsim 2400$\,K planets~\citep{gandhi_retrieval_2023, guo_detection_2024, guo_elemental_2026, sanchez_stellar_2026} appears to occur at higher temperatures and also roughly matches expectations based on the nightside temperature (Figure~\ref{fig:population}, bottom panel). While Ti-species have been reported on TOI-1518b~\citep{simonnin_time-resolved_2025}, WASP-33b~\citep{cont_atmospheric_2022}, TOI-2109b (Pelletier et al.\ in prep), and KELT-9b~\citep{hoeijmakers_atomic_2018}, their abundances remain poorly characterised. Most other known exoplanets have not been observed with adequate observations to be able to detect, or rule out, the presence of Fe- or Ti-species (Figure~\ref{fig:population}). Notably this means that equilibrium chemistry calculations based on local temperature at the terminator or dayside may mispredict the composition of atmospheres that are affected by nightside cold trapping, for example by overpredicting the TiO abundance which can then bias any inferred results~\citep[e.g.][see their Figure 7]{pelletier_enriched_2026}. However, future observations may yet find Fe or Ti to be present and abundant in even colder planets, potentially altering this emerging picture based on relatively few measurements. Indeed, the lack of any atmosphere showing depletions of Fe relative to other metals~\citep{gandhi_retrieval_2023, pelletier_vanadium_2023} suggests that iron is likely present in the gas phase on at least some planets colder than WASP-76b. Contrastingly, the significant underabundance of Ti on WASP-121b and WASP-76b implies that even colder planets will completely lack titanium in the gas phase.

The gradual onset of titanium in ultra-hot Jupiter atmospheres, if confirmed, could mean that nightside cold traps vary in sequestering effectiveness, or that advection and vertical mixing can partly counteract them in some cases. To deplete a species from the photosphere of a gaseous planet, nightside condensation alone is not sufficient as any formed cloud particles would recirculate to the dayside and vapourise back to the gas phase. If condensates can rainout, however, they can sink to the deeper atmosphere where they would not necessarily be observable even upon being recirculated back to the dayside at higher pressures, depending on the vertical mixing.  The fraction of observable titanium would thus depend on the sequestering rate, which is likely proportional to the effective condensation area (i.e.\ how much of the nightside is cold enough to allow cloud formation), and the vertical mixing that can bring rained out condensates back to the hotter upper atmosphere.

As a simple picture, one can think of a planetary nightside as having a given surface area where the temperature is low enough for condensation and rainout to be possible.  In this scenario, the colder WASP-76b would have a large fraction of its nightside where condensation can occur, allowing Ti to be nearly fully removed from the upper atmosphere.  By comparison, the slightly hotter WASP-121b would have a smaller nightside condensation area, and/or more mixing, enabling some Ti to remain in the gas phase at photospheric pressures.  Meanwhile WASP-189b would have an even smaller nightside condensation area, with hotter planets potentially not having any part of their nightsides where rainout is possible. More observations of the hottest known planets such as WASP-33b, TOI-2109b, or even KELT-9b will be useful to map out the limit of the titanium transition region.

Observations spanning a broader range of parameters will be needed to better understand the release of titanium to the gas phase in the atmosphere of hot giant exoplanets, and its role (via TiO) in driving thermal inversions, given that the cold-trapping efficiency likely depends on other planetary properties such as the metallicity and surface gravity, which will affect the settling timescale of forming cloud particles~\citep[e.g.][]{spiegel_can_2009}.  While WASP-189b is denser than WASP-76b and WASP-121b that show stronger titanium depletion levels (Figure~\ref{fig:population}), higher and lower $\log g$ limit cases such as WASP-18b and HAT-P-67b will be interesting to study in the context of cold trapping. Similarly detailed studies exploring the onset of refractory species with different condensation temperatures (e.g.\ S, Na, Fe, Si) will allow us to better understand how nightside cloud formation and rainout shape hot and ultra-hot Jupiter atmospheres.

\section{Conclusions}\label{sec:conclusion}

We observed the dayside atmosphere of WASP-189b using both the HARPS optical and NIRPS near-infrared high resolution spectrographs. We detect emission features of neutral atomic Fe and Ti via cross-correlation and use both a free and a chemical equilibrium retrieval approach to characterise the atmosphere of WASP-189b, testing the sensitivity of any inferred parameters on model assumptions. Using a flexible TP parameterisation, we recover a vertical temperature structure with a strong inversion, consistent with both previous works and GCM models. We find the atmosphere of WASP-189b to be overall slightly metal-poor, with an iron-to-titanium abundance ratio measured to be marginally sub-stellar from both the free ($[$Ti/Fe$]_{*}$ = $-0.43_{-0.30}^{+0.31}$) and chemical equilibrium ($[$Ti/Fe$]_{*}$ = $-0.33_{-0.23}^{+0.24}$) retrievals.

Given that we expect Fe and Ti to be in the same proportion as in the star, any deviation is likely due to atmospheric phenomena, rather than an inherent underabundance of Ti on WASP-189b. As Ti is more refractory than Fe, one possible explanation for its underabundance could be partial nightside cold trapping. In the context of the population, we find that the gradual onset of Ti in ultra-hot Jupiter atmospheres correlates with when Ti is expected to vapourise based on the estimated nightside temperature. This suggests that Ti species will always appear underabundant relative to equilibrium chemistry model predictions if assuming solar or stellar abundance ratios but neglecting nightside cold trapping. This would also imply that TiO may not be the primary driver of thermal inversions as long hypothesised~\citep[e.g.][]{hubeny_possible_2003, fortney_unified_2008}, at least on the colder end of the ultra-hot Jupiter population.  However, in a metal rich atmosphere, there could still be sufficient TiO to cause an inversion, even if it is underabundant relative to other metals.  
Alternatively, other optical absorbers such as H$^{-}$, VO, Fe, and/or a combination of metal oxides, hydrides, and atomic metals could potentially also contribute enough optical/UV opacity to create thermal inversions in ultra-hot Jupiter atmospheres~\citep{spiegel_can_2009, lothringer_extremely_2018, gandhi_new_2019, piette_assessing_2020, petz_pepsi_2025}. In this picture, no gaseous forms of titanium should be detectable in planets with equilibrium temperatures colder than $\sim$2200\,K, unless they have a very low surface gravity.  While trends are beginning to emerge, more detailed studies exploring the ratio of refractory metals with different condensation temperatures will provide crucial insights into how nightside cloud formation and rainout shape the thermal structure and chemistry of exoplanetary atmospheres. 

\vspace{-4mm}

\section*{Data availability}
\vspace{-1mm}
The NIRPS and HARPS data used in this work is available on DACE (\url{https://dace.unige.ch}).

\begin{acknowledgements}
The authors thank the anonymous referee for providing insights that improved the quality of this manuscript. S.P.\ thanks Bibiana Prinoth for general discussions relating to titanium in exoplanet atmospheres. Based on observations taken at the European Southern Observatory for programme 111.2506.001 (PI: F.~Bouchy). This project has been carried out within the framework of the National Centre of Competence in Research PlanetS supported by the Swiss National Science Foundation (SNSF) under grants 51NF40\_182901 and 51NF40\_205606. The authors acknowledge the financial support of the SNSF. RA  acknowledges the SNSF support under the Post-Doc Mobility grant P500PT\_222212 and the support of the Institut Trottier de Recherche sur les Exoplan\`etes (IREx). RA, LDau, LMo, \'EA, FBa, BB, NJC, RD, LMa, CC \& JPW  acknowledge the financial support of the Fonds de Recherche du Qu\'ebec - Secteur Nature et Technologies (FRQ-NT) through the Centre de recherche en astrophysique du Qu\'ebec as well as the support from the Trottier Family Foundation and IREx. DE acknowledges support from the SNSF for project 200021\_200726. EC, SCB, ED-M \& NCS acknowledge the support from FCT - Funda\c{c}\~ao para a Ci\^encia e a Tecnologia through national funds by these grants: UIDB/04434/2020, UIDP/04434/2020. LDau  acknowledges the support of the Natural Sciences and Engineering Research Council of Canada (NSERC) and from the FRQ-NT. LMo  acknowledges the support of the NSERC (reference number 589653). NBC acknowledges support from an NSERC Discovery Grant, a Canada Research Chair, and an Arthur B.\ McDonald Fellowship, and thanks the Trottier Space Institute for its financial support and dynamic intellectual environment. 
TF, XB, XDe \& VY  acknowledge funding from the French ANR under contract number ANR\-24\-CE49\-3397 (ORVET), and the French National Research Agency in the framework of the Investissements d'Avenir program (ANR-15-IDEX-02), through the funding of the ``Origin of Life" project of the Grenoble-Alpes University. \'EA, FBa, RD \& LMa  acknowledges support from Canada Foundation for Innovation (CFI) program, the Universit\'e de Montr\'eal and Universit\'e Laval, the Canada Economic Development (CED) program and the Ministere of Economy, Innovation and Energy (MEIE). SCB acknowledges the support from Funda\c{c}\~ao para a Ci\^encia e Tecnologia (FCT) in the form of a work contract through the Scientific Employment Incentive program with reference 2023.06687.CEECIND. Research activities of the Board of Observational and Instrumental Astronomy at the Federal University of Rio Grande do Norte (NAOS) are supported by continuous grants from the Brazilian funding agency CNPq. This study was financed in part by the Coordena\c{c}\~ao de Aperfei\c{c}oamento de Pessoal de N\'ivel Superior -- Brasil (CAPES) -- Finance Code 001, and by the program CAPES/Print. BLCM acknowledges the CAPES postdoctoral and CNPq research fellowships (Grant No.\ 305804/2022-7). RC  acknowledges support from the Canada Research Chairs Program and the NSERC. JRM  acknowledges CNPq research fellowships (Grant No.\ 308928/2019-9). ED-M  acknowledges the support from FCT through Stimulus FCT contract 2021.01294.CEECIND and by the Ram\'on y Cajal contract RyC2022-035854-I funded by MICIU/AEI/10.13039/501100011033 and by ESF+. JIGH, RR, ASM \& AKS  acknowledge financial support from the Spanish Ministry of Science, Innovation and Universities (MICIU) projects PID2020-117493GB-I00 and PID2023-149982NB-I00. ICL  acknowledges CNPq research fellowships (Grant No.\ 313103/2022-4). CMo acknowledges the funding from the SNSF under grant 200021\_204847 “PlanetsInTime”. Co-funded by the European Union (ERC, FIERCE, 101052347). Views and opinions expressed are however those of the author(s) only and do not necessarily reflect those of the European Union or the European Research Council (ERC). Neither the European Union nor the granting authority can be held responsible for them. GAW is supported by a Discovery Grant from the NSERC of Canada. ED acknowledges support from a Banting Postdoctoral Fellowship - NSERC, the Faculty of Science at the University of Waterloo, and the Waterloo Centre for Astrophysics. XDu acknowledges the support from the ERC under the European Union’s Horizon 2020 research and innovation programme (grant agreement SCORE No 851555) and from the SNSF under the grant SPECTRE (No 200021\_215200). Y.G.C.F.\ acknowledges funding from the SNSF through project P500PN\_217951. FG acknowledges support from the FRQ-NT under file \#350366. KAM acknowledges support from the SNSF Postdoc Mobility grant P500PT\_230225. CP acknowledges support from the NSERC Vanier scholarship, the Trottier Family Foundation, and the E.\ Margaret Burbidge Prize Postdoctoral Fellowship from the Brinson Foundation. AKS acknowledges financial support from La Caixa Foundation (ID 100010434) under the grant LCF/BQ/DI23/11990071. This publication makes use of The Data \& Analysis Center for Exoplanets (DACE), which is a facility based at the University of Geneva (CH) dedicated to extrasolar planets data visualization, exchange and analysis. DACE is a platform of the Swiss NCCR PlanetS, federating the Swiss expertise in Exoplanet research.

\end{acknowledgements}

\vspace{-8mm}

\bibliographystyle{aa}
\bibliography{W189_NIRPS}

\clearpage

\begin{appendix}

\onecolumn
\nolinenumbers

\section{Cross-correlation non-detections}\label{appendix:non_detections}
\begin{figure*}[ht]
\begin{center}
\includegraphics[width=\linewidth]{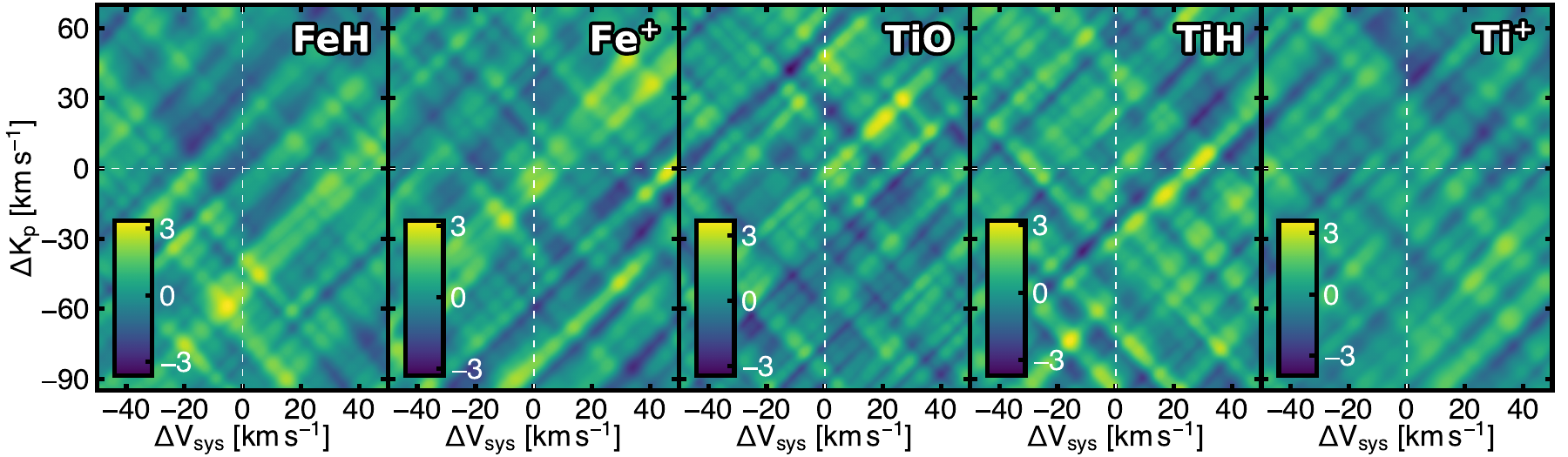}
\end{center}
\vspace{-5mm}
\caption{
Same as Figure~\ref{fig:CCF_maps} bottom panels but for chemical species for which no significant signal is observed.  We note that a non-detection from these data does not imply that a species is not present in the gas phase on the dayside atmosphere of WASP-189b, but rather can be due to a lack of sensitivity.  
}
\label{fig:CCF_non_detections}
\end{figure*}

Here we shown cross-correlation signal-to-noise maps for Fe- and Ti-bearing species for which no clear signal was found (Figure~\ref{fig:CCF_non_detections}).  We caution, however, that thermal emission observations are not as sensitive as transit spectroscopy for ions which mostly have spectral features at bluer wavelengths where the planet-to-star flux ratio is small. As such, our non-detection of Fe$^+$ and Ti$^+$ does not mean that these ions are non present on the dayside atmosphere of WASP-189b, simply that these data are not sensitive to their spectral contributions.  Indeed even high abundances of Fe$^+$ and Ti$^+$ cannot be ruled out by these data (e.g.\ Figure~\ref{fig:TP_abun}, right panel).

\clearpage

\section{Sensitivity of velocity offsets}\label{appendix:KpVsys_tests}

We explore possible sources of bias in velocity offsets of observed exoplanetary signals observed with high resolution spectroscopy.  With ultra-hot Jupiters often accelerating by a few kilometres per second every minute, small timing or ephemeris errors can translate into offsets of kilometres per second, which could be misinterpreted as evidence of winds or 3D effects.  In particular, exposure header information provided by the data reduction pipeline of instruments often give the exposure start time in JD.  To precisely compute the expected planet orbital velocities, one should use the (ideally flux-weighted) mid-exposure time in the frame invariant Barycentric Dynamical Time (BJD$_{\mathrm{TDB}}$) to account for finite light travel time and general relativistic effects~\citep{eastman_achieving_2010}. It is also important to verify that the reference $T_0$ is the most precise latest available and is also in BJD$_{\mathrm{TDB}}$ as papers quoting $T_0$ values will differ in convention. For example, \cite{lendl_hot_2020} provide the $T_0$ in BJD$_{\mathrm{TT}}$, \cite{ivshina_tess_2022} give values in BJD$_{\mathrm{TDB}}$, and \cite{deline_atmosphere_2022} use BJD without specifying which type (although the authors confirmed that this is indeed BJD$_{\mathrm{TDB}}$, priv.\ comm.).  Even if consistently using JD/UTC values for both $T_0$ and the observations, timing differences of several minutes can still be incurred considering that light takes about 16.6 minutes to transverse Earth's orbit.  Especially for short-period planets, this can then translate into kilometres per second offsets in observed signals, which could be misinterpreted as dynamical and/or 3D effects.

\begin{figure*}[ht]
\begin{center}
\includegraphics[width=\linewidth]{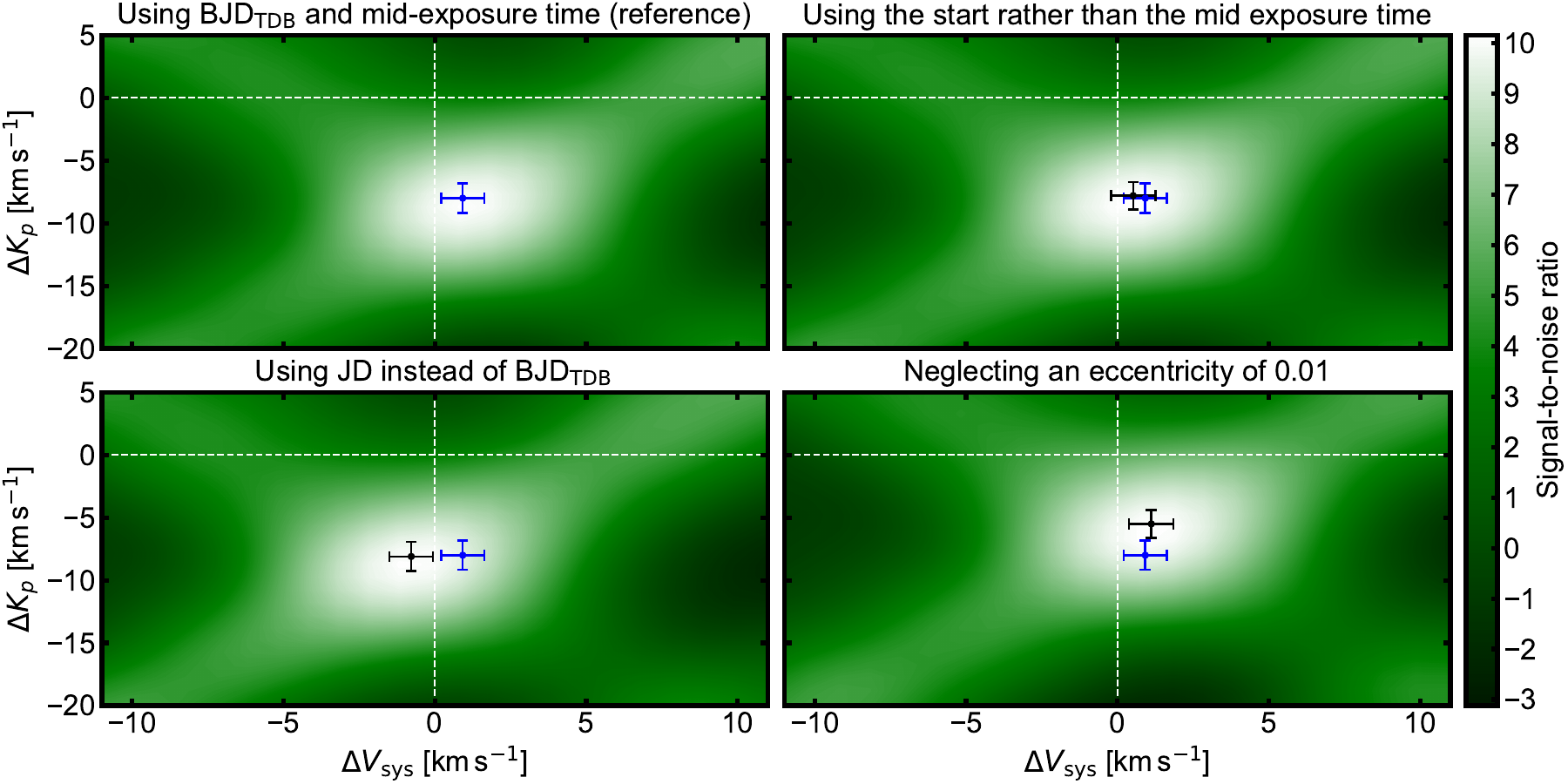}
\end{center}
\vspace{-5mm}
\caption{Same as Figure~\ref{fig:CCF_maps} bottom left panel but under different ephemeris assumptions, zoomed in on the planetary signal.  The top left panel is the approach used in our analysis (reference), computing the orbital velocities of WASP-189b using the Barycentric Dynamical Time (BJD$_{\mathrm{TDB}}$) at mid-exposure as the time stamp of every exposure.  
The top right panel instead uses the exposure start time as the time stamp of each exposure, which results in a timing offset of $t_{\mathrm{exp}}/2$ that can be important especially for long exposures of short period planets.  
The bottom left panel erroneously uses the Julian Date (JD) time rather than the BJD$_{\mathrm{TDB}}$, which can result in a few kilometres per second offset in $V_{\mathrm{sys}}$.
The bottom right panel computes the expected orbital velocities of WASP-189b assuming an eccentricity of 0.01 (as opposed to assuming a circular orbit as in the reference case), which can bias the observed $K_p$. Using accurate timings is critical for interpreting velocity offsets in the context of dynamics and 3D effects.
}
\label{fig:CCF_offsets}
\end{figure*}

The importance of accurate timings will mostly affect the inferred $V_\mathrm{sys}$ (and hence the derived $V_\mathrm{wind}$) of the planetary signal (Figure~\ref{fig:CCF_offsets}) and can vary for different observational setups depending on the relative Earth and target positions relative to the barycentre, and the exposure duration.  Furthermore, even a slight eccentricity of 0.01 can results in velocity offsets of a few kilometres per second for $K_p$ (Figure~\ref{fig:CCF_offsets}, bottom right panel, see also \cite{savel_no_2022, meziani_effect_2025}).

\clearpage

\section{Atmospheric retrieval parameters and example corner plot}\label{appendix:corner_plots}

\begin{table*}[h]
\caption{Atmospheric retrieval parameter priors and posteriors.} 
\vspace{-3mm}
\label{tab:retrieval_results}
\centering
\def\arraystretch{1.3}
\begin{tabular}{ccccc}
\hline
\hline
Parameter & Description & Prior & Free & Chemical equilibrium~ \\
\hline
\hline
$\log \chi_{\mathrm{Fe}}$ & $\log_{10}$ abundance of Fe & $\mathcal{U}$[$-12$,$-0.3$] & $-4.68_{-0.31}^{+0.27}$ & --- \\
$\log \chi_{\mathrm{Fe}^+}$ & $\log_{10}$ abundance of Fe$^+$ & $\mathcal{U}$[$-12$,$-0.3$] & $<-$4.77 (2$\sigma$) & --- \\
$\log \chi_{\mathrm{Ti}}$ & $\log_{10}$ abundance of Ti & $\mathcal{U}$[$-12$,$-0.3$] & $-7.91_{-0.29}^{+0.31}$ & --- \\
$\log \chi_{\mathrm{Ti}^+}$ & $\log_{10}$ abundance of Ti$^+$ & $\mathcal{U}$[$-12$,$-0.3$] & $<-$5.65 (2$\sigma$) & --- \\
$\log \chi_{\mathrm{TiO}}$ & $\log_{10}$ abundance of TiO & $\mathcal{U}$[$-12$,$-0.3$] & $-8.54_{-0.24}^{+0.20}$ & --- \\
$\log \chi_{\mathrm{H}^{-}}$ & $\log_{10}$ abundance of H$^-$ & $\mathcal{U}$[$-12$,$-0.3$] & $-9.27 \pm 0.10$ &  --- \\
$\log \chi_{\mathrm{e}^{-}}$ & $\log_{10}$ abundance of $e^-$  & $\mathcal{U}$[$-12$,$-0.3$] & $<-$2.62 (2$\sigma$) &  --- \\
$[$Fe/H$]_{\odot}$ & $\log_{10}$ Fe metallicity rel.\ to solar & $\mathcal{U}$[$-3$,$3$] & --- & $-0.44_{-0.16}^{+0.15}$ \\
$[$Ti/H$]_{\odot}$ & $\log_{10}$ Ti metallicity rel.\ to solar  & $\mathcal{U}$[$-3$,$3$] & --- & $-0.94_{-0.11}^{+0.12}$ \\
T$_{1\,\mu \mathrm{bar}}$ & Temperature at $10^{-6}\,\mathrm{bar}$ (K) & $\mathcal{U}$[100,8000] &  $3032_{-1516}^{+1378}$  & $2834_{-1714}^{+1828}$ \\
T$_{10\,\mu \mathrm{bar}}$ & Temperature at $10^{-5}\,\mathrm{bar}$ (K) & $\mathcal{U}$[100,8000] &  $3359_{-740}^{+551}$  & $3240_{-836}^{+878}$ \\
T$_{0.1\,\mathrm{mbar}}$ & Temperature at $10^{-4}\,\mathrm{bar}$ (K) & $\mathcal{U}$[100,8000] &  $3991_{-228}^{+212}$  & $3738_{-336}^{+358}$ \\
T$_{1\,\mathrm{mbar}}$  & Temperature at $10^{-3}\,\mathrm{bar}$ (K) & $\mathcal{U}$[100,8000] &  $3940_{-291}^{+276}$  & $4036_{-175}^{+178}$ \\
T$_{10\,\mathrm{mbar}}$ & Temperature at $10^{-2}\,\mathrm{bar}$ (K) & $\mathcal{U}$[100,8000] &  $2487_{-332}^{+305}$  & $3145_{-174}^{+140}$ \\
T$_{0.1\,\mathrm{bar}}$ & Temperature at $10^{-1}\,\mathrm{bar}$ (K) & $\mathcal{U}$[100,8000] &  $2009_{-312}^{+303}$  & $2163_{-224}^{+194}$ \\
T$_{1\,\mathrm{bar}}$ & Temperature at $10^{0}\,\mathrm{bar}$ (K) & $\mathcal{U}$[100,8000] &  $2975_{-196}^{+143}$  & $2967_{-213}^{+173}$ \\
T$_{10\,\mathrm{bar}}$ & Temperature at $10^{1}\,\mathrm{bar}$ (K) & $\mathcal{U}$[100,8000] &  $3467_{-1037}^{+1120}$  & $3538_{-703}^{+674}$ \\
$K_p$ & Keplerian velocity (km\,s$^{-1}$)  & $\mathcal{U}$[$-150$,$250$] &   $192.2 \pm 0.6$  & $192.8_{-0.7}^{+0.8}$ \\
$V_\mathrm{sys}$ & Systemic velocity (km\,s$^{-1}$)  & $\mathcal{U}$[$-70$,$30$] &  $-21.0 \pm 0.3$ & $-20.9 \pm 0.4$ \\
\hline
$[$Ti/H$]_{\odot} - [$Fe/H$]_{\odot}$ & $\log_{10}$ Ti/Fe rel.\ to solar & --- & $-0.62\pm0.30$ & $-0.51_{-0.22}^{+0.23}$  \\
$[$Ti/H$]_{*} - [$Fe/H$]_{*}$ & $\log_{10}$ Ti/Fe rel.\ to stellar & --- & $-0.43_{-0.30}^{+0.31}$ & $-0.33_{-0.23}^{+0.24}$ \\
\hline
\end{tabular}\\ 
\vspace{-1mm}
\end{table*}

\begin{figure*}
\begin{center}
\includegraphics[width=\linewidth]{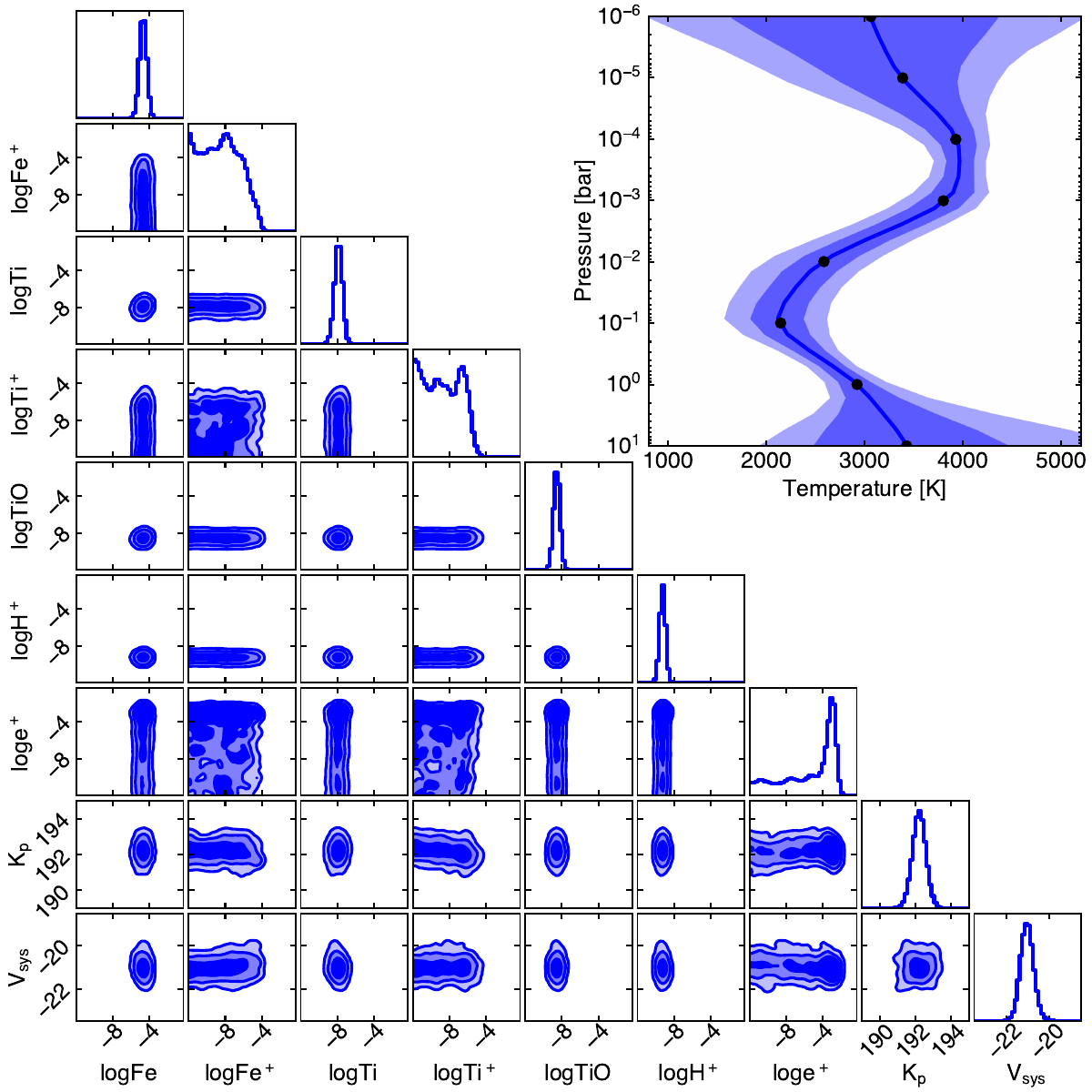}
\end{center}
\vspace{-5mm}
\caption{
Corner plot of the marginalized posterior distributions of the atmospheric and orbital properties of WASP-189b obtained from the free retrieval analysis. Shaded regions respectively depict 39.3\%, 86.5\%, and 98.9\% confidence intervals. The recovered temperature structure from the eight temperature points (not shown in the corner plot for clarity) is shown in the top right panel.
}
\label{fig:corner_free}
\end{figure*}

\end{appendix}

\end{document}